# Signatures of Strong Magnetization and Metal-Poor Atmosphere for a Neptune-Size Exoplanet

Lotfi Ben-Jaffel[1,2*], Gilda E. Ballester[2], Antonio García Muñoz [3,4], Panayotis Lavvas[5], David K. Sing[6,7], Jorge Sanz-Forcada[8], Ofer Cohen[9], Tiffany Kataria[10], Gregory W. Henry[11], Lars Buchhave[12], Thomas Mikal-Evans[13], Hannah R. Wakeford[14], Mercedes López-Morales[15]

[1]*Institut d'Astrophysique de Paris. Sorbonne Universités, UPMC & CNRS, 98 bis Bd. Arago, F-75014 Paris, France.*
[2]*Lunar and Planetary Laboratory, University of Arizona, 1541 E Univ. Blvd., Tucson, Arizona 85721, USA.*
[3]*Zentrum für Astronomie und Astrophysik, Technische Universität Berlin, D-10623 Berlin, Germany.*
[4]*AIM, CEA, CNRS, Université Paris-Saclay, Université de Paris, F-91191 Gif-sur-Yvette, France.*
[5]*Groupe de Spectroscopie Moléculaire et Atmosphérique, Université de Reims, Champagne-Ardenne, CNRS UMR 7331, France.*
[6]*Department of Earth & Planetary Sciences, John Hopkins University, Baltimore, MD, USA*
[7]*Department of Physics & Astronomy, John Hopkins University, Baltimore, MD, USA*
[8]*Centro de Astrobiología (CSIC-INTA), ESAC, P.O. Box 78, E-28691, Villanueva de la Cañada, Madrid, Spain.*
[9]*Lowell Center for Space Science and Technology, University of Massachusetts, Lowell, MA 01854, USA.*
[10]*NASA Jet Propulsion Laboratory, 4800 Oak Grove Drive, Pasadena, CA 91109, USA.*
[11]*Center of Excellence in Information Systems, Tennessee State University, Nashville, TN 37209, USA.*
[12]*DTU Space, National Space Institute, Technical University of Denmark, Elektrovej 328, DK-2800 Kgs. Lyngby, Denmark*
[13]*Kavli Institute for Astrophysics and Space Research, Massachusetts Institute of Technology, Cambridge, MA 02139, USA.*
[14]*School of Physics, University of Bristol, HH Wills Physics Laboratory, Tyndall Avenue, Bristol 1TL, UK.*
[15]*Center for Astrophysics | Harvard & Smithsonian, 60 Garden Street, Cambridge, MA 02138, USA.*
*e-mail: bjaffel@iap.fr*

**The magnetosphere of an exoplanet has yet to be unambiguously detected. Investigations of star-planet interaction and neutral atomic hydrogen absorption during transit to detect magnetic fields in hot Jupiters have been inconclusive, and interpretations of the transit absorption non-unique. In contrast, ionized species escaping a magnetized exoplanet, particularly from the polar caps, should populate the magnetosphere, allowing detection of different regions from the plasmasphere to the extended magnetotail, and characterization of the magnetic field producing them. Here, we report ultraviolet observations of HAT-P-11b, a low-mass (0.08 $M_J$) exoplanet showing strong, phase-extended transit absorption of neutral hydrogen (maximum and tail transit depths of 32 ± 4%, 27 ± 4%) and singly ionized carbon (15 ± 4%, 12.5 ± 4%). We show that the atmosphere should have less than six times the solar metallicity (at 200 bars), and the exoplanet must also have an extended magnetotail (1.8–3.1 AU). The HAT-P-11b equatorial magnetic field strength should be about 1–5 Gauss. Our panchromatic approach using ionized species to simultaneously derive metallicity and magnetic field strength can now constrain interior and dynamo models of exoplanets, with implications for formation and evolution scenarios.**

Lines of hydrogen, helium, carbon, oxygen and heavy metals have been successfully detected in the UV, optical and near-IR on a few exoplanets with significant absorption extending beyond their Roche lobe[1-6]. Those detections are consistent with the immense X-ray/extreme-UV irradiation from the host star that drives hydrodynamic escape from microbar pressure levels. For magnetized exoplanets, the interaction between the stellar wind and the planetary magnetic field produces large-scale magnetospheric structure (Fig. 1, 4). The planetary wind can fill-in the planet's plasmasphere[7] (inner magnetospheric region) and magnetotail (far region on the night side), and thus extend the gas distribution blocking the starlight during transit. Unfortunately, we have not yet been able to relate the transit absorption by the outer layers to the underlying magnetospheric structure nor to the bulk composition in the deep atmosphere.

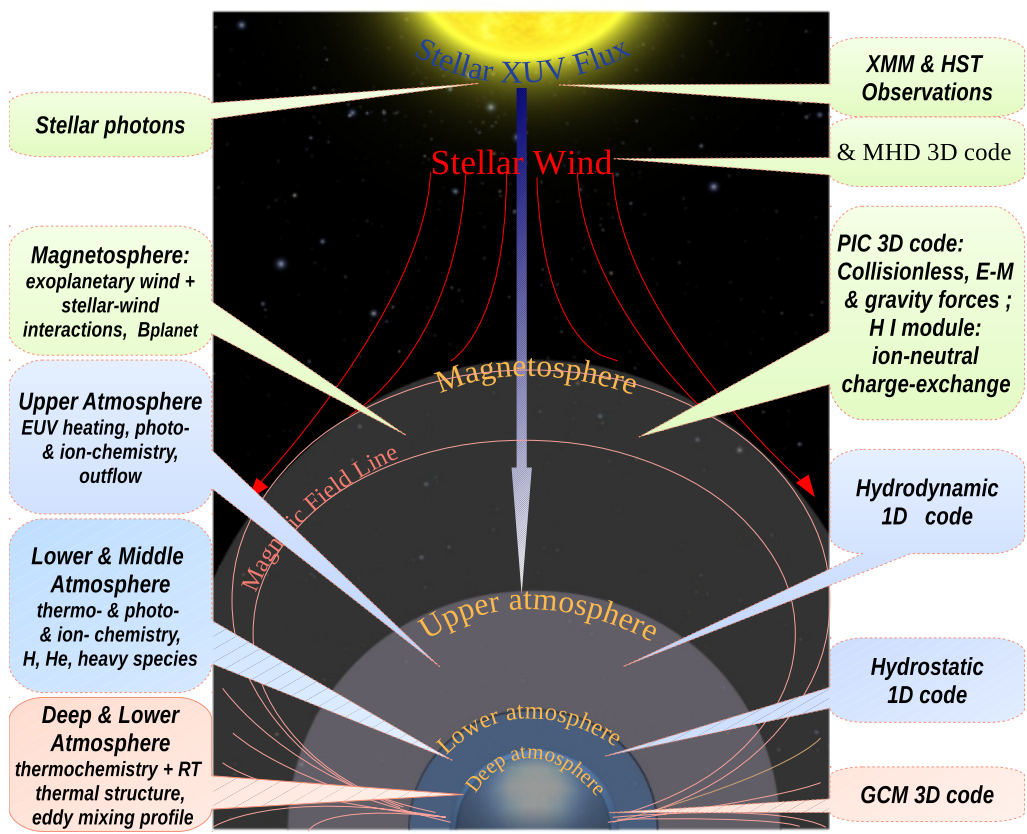

**Figure 1**: **Flowchart of the HAT-P-11 star-planet system (not at scale) and the corresponding modelling framework proposed in this study**. Our panchromatic comprehensive approach consists in modelling the different interconnected atmospheric layers and the magnetosphere that are powered by the stellar flux (large arrow) and the stellar wind (red). For each layer, the main processes are described on left and the associated model tools are shown on the right. Few magnetic field lines are shown (pink). EUV (extreme ultraviolet) and RT (radiation transfer). The framework is described in details in Methods, section I.

For instance, the composition, the magnetisation level and their time evolution over the lifetime of exoplanets are open questions. Despite the few thousand exoplanets known to date, the quest for detecting intrinsic exoplanetary magnetic activity remains unsuccessful. Based on solar system magnetism, weak radio signals are expected from a subset of exoplanets, a signal that existing technology has either failed to detect or to unambiguously associate with an exoplanet[8]. Star-planet interactions, detected as a planet-modulation of the stellar chromospheric emission in Ca II by some hot Jupiters, have also been invoked to constrain the magnetic field strength. The method is indirect and relates to specific exoplanet/star combinations (magnetic field configuration producing magnetic reconnection, size of the exoplanet, orbital distance relative to the star's Alfven radius, etc.)[9,10]. Strong transit absorption in neutral atomic hydrogen has also been associated, via charge exchange with stellar protons, with the exoplanet magnetospheric cavity, yet the link with the magnetic field is controversial and the interpretation of the transit signature is not unique[11]. Here, we use a novel approach to implement panchromatic observations of the HAT-P-11 system plus interrelated simulations and corresponding uncertainties carefully linking the physical conditions in the deep exoplanet atmosphere (200 bar) to all the atmospheric layers above it, up to the magnetosphere (Figure 1 and Methods). From detailed comparison with the observations, we uncover the presence of a plasmasphere and a magnetotail, constrain the atmospheric metallicity compared to existing gas planets in our solar system, and estimate the intrinsic magnetic field of the exoplanet (Methods and Extended Data Figs. 1-7).

HAT-P-11b is a warm, low mass (equilibrium temperature $T_{eq} \sim 870K$) exoplanet orbiting an active K4 main sequence star at $\sim 0.0465$ AU[12,13]. HAT-P-11b is among the few low-mass exoplanets showing water in its lower atmosphere[14,15]. He I absorption has also been detected in its upper atmosphere[16]. An extensive *Kepler* dataset showed HAT-P-11b to have an approximately polar and eccentric orbit, evidencing a dynamically disturbed history for the system[17,18]. Recently, a second, non-transiting, Jupiter-mass exoplanet was found in the system, with an eccentric and tilted orbit[19].

**Hubble Space Telescope (*HST*) Observations:**
We performed several transit observations in the far-UV under the *Hubble Space Telescope (HST)* Panchromatic Comparative Exoplanetary Treasury program (PanCET). We observed four transits of HAT-P-11b on October 28 2016, December 16 & 21 2016, and May 21 2017 using the Cosmic Origins Spectrograph (COS) with the G130M grating, sampling the far-UV ~113-146 nm spectral region at medium resolution (~15 km/s). We also observed two transits on October 23 and November 12 2016 using the Space Telescope Imaging Spectrograph (STIS) with the G140M grating, covering ~119.4-124.9 nm at medium resolution with ~12.3 km/s/pixel dispersion. The neutral oxygen O I 130.4 nm triplet and the ionized carbon C II 133.5 nm doublet were observed with COS, while STIS observed the H I Lyman-α (Lyα) line at 121.567nm.

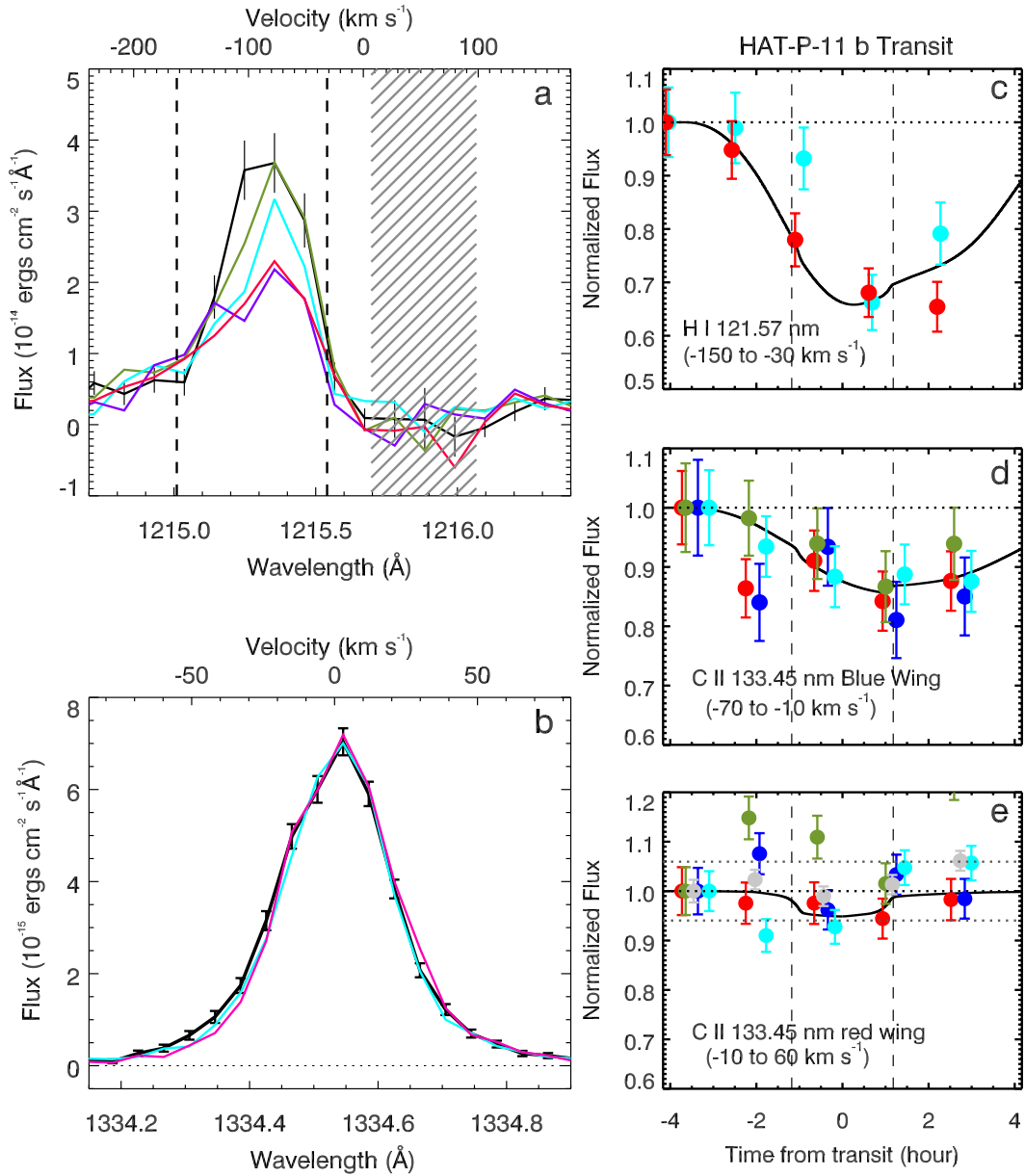

**Figure 2**: **HAT-P-11b FUV transit absorption.** Error bars represent the 1σ statistical uncertainties that have been propagated from STIS and COS data reduction pipeline. **a,b,** Stellar H I 121.57 nm (**a**) and C II 133.45 nm (**b**) line absorption by the exoplanet during transit. **c-e,** Transit absorption versus exoplanet orbital phase. For each visit, fluxes are normalized by the flux of the first orbit occurring before the transit event. Horizontal dotted line lines at normalized flux = 1 represent the absorption reference level. The optical transit duration is indicated by vertical dashed lines. One of our best fits (B~2.4 G) to both C II and H I lines are shown (solid) for the same model. **a,** H I 121.57 Lyα line profile binned by two in wavelength for the five orbital phases indicated in Figure 2c with respect to the optical transit: out-of-transit phase 1 (black), pre-transit phase 2 (olive green), ingress phase 3 (turquoise), in-transit phase 4 (red), and post-transit phase 5 (purple). The plotted line shapes are from the second STIS visit. The hatched area indicates the spectral window of the sky background contamination. **b,** C II 133.45 nm line profile binned by 4 in wavelength for the three orbital phases indicated in Figure 2d: out-of-transit phase 1 (black), average of in-transit phases 3 & 4 (magenta), and post-transit phase 5 (turquoise). The deviant pixel on the red wing of the magenta line profile is a single statistical effect (see Fig. 2e). The plotted line shapes are from the average of all four COS visits. **c,** H I Lya line flux integrated over projected velocities from -150 km/s to -30 km/s of the absorbing H I atoms (indicated by two dashed vertical lines in Figure 2a) for two transit events: HST visit 1 on 10/23/2016 (turquoise) and HST visit 2 on 11/11/2016 (red). **d,** C II 133.45 nm line flux integrated over the blue wing for projected velocities from -70 km/s to -10 km/s for the four HST transit events: transit 1 (red), transit 2 (olive), transit 3 (blue), and transit 4 (aqua). **e,** Same as **d**, but for the red wing of the C II line integrated from -10 km/s to +50 km/s. Horizontal dotted lines are shown at ±3σ from the normalized flux = 1. We also show the average of the four COS visits (grey).

Comparing stellar spectra before transit to that during transit we find increasing H I Lya absorption with HAT-P-11b's orbital phase (Fig.2a). Most of the absorption occurs at velocities between -150 km/s to -30 km/s in the reference frame of the star. When averaged over available transits, the integrated blue wing of the stellar Lya line shows pre-ingress absorption of 14± 5% (1σ) at the time of the optical-disk ingress, goes as deep as 32.5 ± 4.5% around mid-transit, and expands far (2.25 h) and deep (28 ± 5%) during egress (Fig. 2b). The corresponding detection levels are ~3σ, ~7σ, and ~5.5σ respectively for ingress, transit, and egress.

To check for stellar variability during each HST observation, we monitor the target's signal every 30 seconds using the time tag information stored in the data. This is achieved for every single stellar emission line, providing a timeseries of the corresponding integrated flux over the exposure time. The analysis of those timeseries showed the absence of statistically significant stellar signal variability above noise levels during orbits 1 & 2 in both visits, respectively ~4 h and ~2.5 h before the optical mid-transit, which supports that the spectra obtained during orbit 1 provide a true out-of-transit stellar reference.

The repeatability of the Lya transit light curve in the two STIS visits and the absence of apparent transit absorption in other stellar transition-region emission lines, observed with the same grating, further indicate that the Lya detection is not due to stellar activity or known instrumental effects (see Methods, section II).

For the C II 133.45 nm line, the comparison of the line profile during transit and egress to the off-transit reference in orbit-1 clearly shows blue Doppler-shifted absorption, particularly over the -70 km/s to -10 km/s spectral range. In contrast, no significant absorption is detected at line center or in the red wing (Fig. 2e). Independently of any modelling, we find that these two spectral signatures strongly constrain the physics of the escaping atoms. The light curve of the integrated blue-wing absorption shows a transit depth of 10± 4% at ~2 hr before the optical ingress, goes as deep as 15 ± 4% at mid-transit and expands far (2.25 hr) and deep (12.5 ± 4%) during egress (Fig. 2d). In contrast, the red wing of the same stellar line shows a flat trend (Fig. 2e). Together, the absence of variability in the stellar flux during the first orbit in all visits, the repeatability of the C II 133.45 nm blue-wing absorption with planet orbital phase for all four COS visits, and the absence of similar transit absorption in other stellar lines (particularly in the red wing of the same C II 133.45 nm line), all indicate that the C II detection is not due to stellar variations or any known instrumental effect (see Methods, section II).

Remarkably, the C II 133.45 nm and Lyα transit light curves look similar, yet the absorption is stronger in Lyα. In addition, the C II 133.45 nm transit absorption is maximum around -50 km/s from line center, while the blue side of Lyα the absorption is maximum around -100 km/s, indicating a factor of two higher average Doppler velocities for H I than C II. The lack of detectable absorption at the center and red wing of the C II 133.4 nm line supports a projected Doppler-shifted absorption process during transit and thus, a dominant global particle motion away from the star. This diagnostic is possible because we can study the entire C II stellar line, in contrast to the wider H I Lyα line that is strongly affected by the ISM.

**Transit absorption: a comprehensive approach**:
Abundances and Doppler velocities inferred from transmission spectroscopy in the FUV are sensitive to the stellar's line width, the cross section of the escaping atom, its abundance and velocity, and its ionization lifetime. Because the system is complex, there is no straightforward one-to-one relationship between a property of the system and a specific aspect of the observed signal. Only a forward comprehensive modeling is able to disentangle the problem (see Methods, section I-IV).

**Free parameters: a limited number for a tractable analysis**
In this comprehensive study, we consider three main free parameters: the metallicity of the deep atmosphere in the range 1-, 2-, 6-, 10-, 30-, 50-, 100- and 150-times solar, the strength of the exoplanet magnetic field in the range ~0.12, 1.2, 2.4, 4.8, and 9.6 G, and the full length of the magnetotail. The carbon abundance is tightly dependent on the deep atmosphere (200 bar) metallicity while the spatial extent of the ionized CII species, which dominate the upper atmosphere, is controlled by the strength of the planetary magnetic field and the induced currents in the inner and outer regions of the magnetosphere. We also consider the tilt of the planet magnetic field axis with respect to its spin axis as a free parameter but within a limited range as required by the fitting process. The reasoning behind the selected ranges is explained in Supplementary Discussion I.

To model and interpret the *HST* transit observations of HAT-P-11b, we connect the deep composition of the exoplanetary atmosphere directly with the planet's atmospheric escape, consistently with the energy input from the stellar radiation and wind (e.g., Fig. 1 and Methods, section III).

The implementation of the magnetospheric modelling follows four simple steps that aim to disentangle the contribution of metallicity versus field strength, while fitting both the C II and H I transit absorptions.

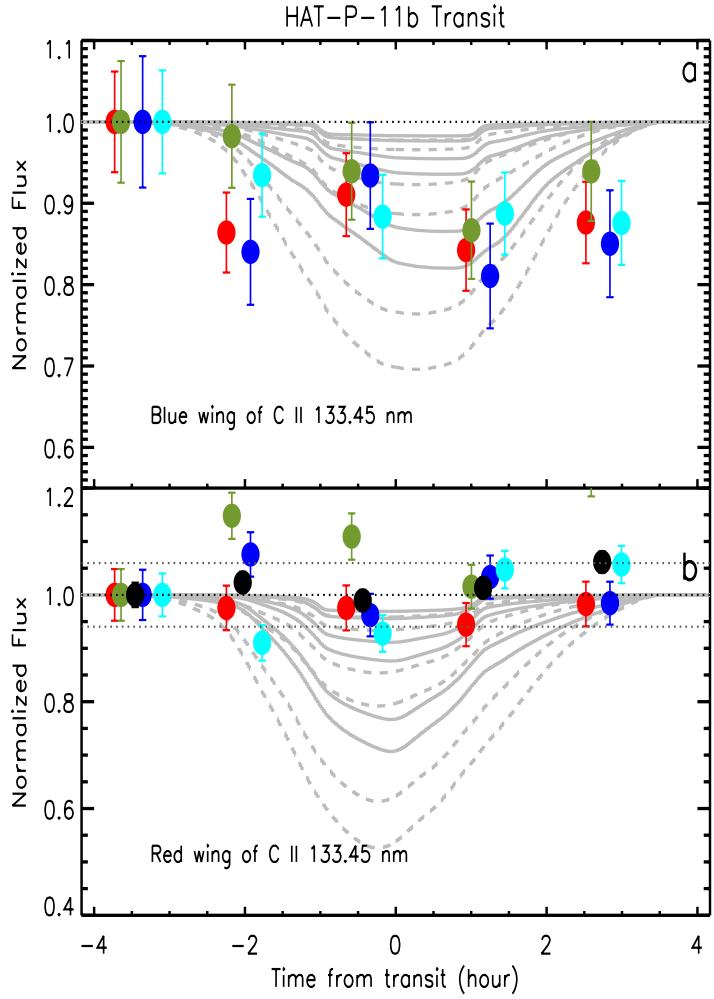

**Figure 3: Sensitivity of CII model transit light curves versus the metallicity assumed in the deep atmosphere and the strength of the intrinsic magnetic field of the exoplanet.** In all panels, grey light curves from top to bottom correspond to 1-, 2-, 6-, 10-, 30- and 50-times solar metallicity; B=1.2 G (dash light curves) and B=2.4 G (solid). Data points of individual HST visits/orbits and related 1σ statistical error bars are also shown with same colours used in Figure 2. Here, the model considers only the effect of the plasmasphere excluding the extended magnetotail. The final fit to observations, including the whole system, is discussed in Step 2 and Extended data Figure 1, with related $\chi^2$ values provided in Table 1 and an example best-fit found in Fig. 2. **a**, Blue wing of the C II 133.45 nm line. The asymmetry of the light curves for B=2.4 G is remarkable compared to the more symmetric ones obtained for B = 1.2 G. **b**, Red wing of same C II 133.45 nm line. No significant absorption can be measured on the red wing. The average of the four COS visits is show in black. Horizontal (dotted) lines show ±3 σ limits of the combined COS visits. Any model light curve that exceeds those limits is rejected for the assumed B strength.

## Step 1: Metallicity effect and related B field

With O I, C II is the most sensitive species to the atmospheric metallicity. For all magnetic field cases, we generated light curves versus atmospheric metallicity for both the blue and red wings of the 133.45 nm line for comparison with observations (Figure 3).

A first result is that models with metallicity above six times solar (3 times stellar) are more than 3σ away from the average red-wing observations and are thus rejected (Table 1). Our metallicity upper limit is insensitive to the field strength or asymmetry considered. For reference, the 3-sigma constraint on the red-wing transit (~6% absorption, with less than 1% chance to happen) can be understood as a noise upper limit absorption that, if it exists, would spectrally occur in the rest frame of the planet (in order to satisfy the red-wing spectral window), and would be principally produced by thermal atoms magnetically trapped in the dense plasmasphere (Figure 4a). This result is the first strong implication from the asymmetry observed in the C II 133.45 nm line absorption level and transit light curve behaviour.

Our derived solar-like upper limit metallicity of HAT-P-11b confirm the recent finding of low metallicity obtained from optical/IR HST observations[15]. Based on a quite different wavelength window, our independent analysis obtained a more stringent metallicity upper limit of 6 times solar at the 3-σ level detection. At the 1-σ detection level regarding the red-wing signal of the CII 133.45 line, our upper limit should not exceed the stellar metallicity (2 times solar) for HAT-P-11b (e.g., Table 1). With its solar-like metallicity, HAT-P-11b appears as a new kind of low-mass Jovian-like exoplanet, and not a Neptune-like one.

**Step 2: The need for a magnetotail**
For the few-times solar metallicity derived above, and all B cases considered, we find that the C II blue wing models show less absorption than expected, particularly during egress orbit 5. To increase the system opacity, the only option is to extend the magnetotail size. Figure 4 shows how the magnetosphere is composed principally of a dense plasmasphere and a tenuous extended magnetotail. For perspective, magnetotails as long as a few thousand planetary radii (or few AU) have been predicted and observed for planets of the solar system[20,21] (see Supplementary Discussion II).

In the present case, no simulation with the regular spatial grid used in our PIC code can handle such an elongated system. However, all PIC models generated in our study show a steady state plasma flow leaving the system in the downtail sector (e.g., Figure 4c,d). Another significant feature of the magnetotail model outflow is that the planetary polar wind is its dominant source (Supplementary Discussion III). This PIC result is consistent with the low-energy plasma particles recently explored by the four Cluster II spacecraft as the dominant population in the Earth magnetosphere[7,22].

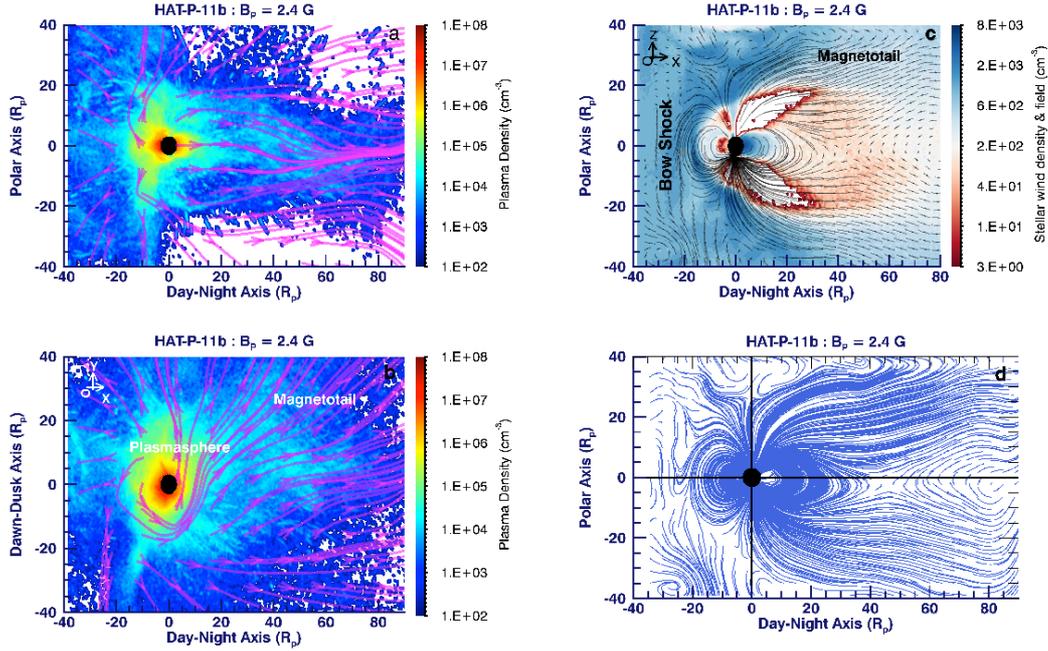

**Figure 4**: **PIC high-resolution simulations (Δ~0.33 $R_p$) for one of the best fits of the magnetic field strength B=2.4 G.** The planetary disk (3 $R_p$) is placed at (0,0) (shown in black). The stellar wind arrives from the left of the figure. Day-night cross sections of the magnetosphere are shown. The reference frame axes are shown on left upper corner in panel **c**. **a**, Planetary protons flow streamlines with arrows (magenta) are shown over the plasma density. The outflow is tailward (to the right). On the dayside, low density and cross-field flow (dark blue) is stellar-ward for lower latitudes but bend tailward at higher latitudes. **b**, Same as in (a) but in the equatorial plane (XY). Ingress corresponds to the top of figure and egress to the bottom (after the tail tilt, provided in Table 1, is applied). We remark the impact of the (anti-clockwise) corotation on the streamlines of the flow directed downtail, confirming early predictions made for the dynamic of the plasma flow due to corotation and an inner source (Io & ionosphere) in the Jovian magnetosphere[26]. **c**, Plasma density of the stellar wind component and the compressed dipole magnetic field lines in the XZ plane. We clearly see an upstream bow shock of the energetic stellar wind (e.g., at around X = -20 Rp, Z=0 Rp) on the dayside. Field aligned particles flow from the near tail to the planetary poles (responsible for planetary aurorae). **d**, Magnetic field lines. We use the separation between open and closed field lines to define the solid angles of the north and south polar caps. A magnetic-field reconnection appears at a distance ~50 $R_P$ on the far magnetotail, which should reinforce downtail plasma ejection from the system over time (and auroral particle precipitation onto the planet).

In this frame, for each case of magnetic field strength, we ran the PIC simulation and used the results to derive the angular extension of the polar cap by finding the separation between open and closed field magnetic field lines. In a second run, the particles originating from the polar cap were then tracked separately from the particles originating at lower latitudes (Figure 5).

To derive the average plasma properties in the magnetotail, we first need to define the spatial size of the magnetopause (MP) inside which the plasma is confined, particularly to account for the tail flattening expected in the OZ direction (see Supplementary Discussion IV). For instance, our PIC simulations show a complex topology of the plasma distribution in the magnetotail that can hardly be described only by tail flattening or averaged kinetic properties (e.g., Figures 4, 5). For that reason, we provide a volume filling factor that should measure the filling/emptiness of the tail region along with a dispersion on most derived average plasma properties for each B case (e.g., Figures 4, 5). This allowed us to derive the average statistical properties of the plasma inside the magnetotail volume (average bulk velocity of the flow), the dispersion on the velocity, and the mean plasma density of each species, and also the related dispersion and the volume filling factor versus the B field strength or tilt (e.g., Table 1).

With the far tail properties derived above, the best fits, in terms of the extended tail size and orientation, are shown in Table 1. First, we find that for largest field strength (8 times the reference value), the ~6-19 AU length of the tail, required to fit the data, is rejected because it would be difficult to physically maintain a cohesive tail structure over such long distances. This case is thus rejected. All other field-strength/metallicity solutions with tail sizes above a few AU are also rejected (Table 1).

**Step 3: the spectral shape of the absorption**
In the steps above, we focused on the transit *integrated* absorption over the blue versus red wings of the C II 133.45 nm line. We now use the spectral profile of the transit absorption. For C II, the signal to noise of all merged data is not as high as for the H I Lya transit, so the comparison at the spectral pixel resolution is not conclusive. In addition, because the transit absorption is affecting the waning slope of a sharp line profile, it is difficult to compare models to observations with relatively large statistical errors. To clarify the diagnostic, we decided to focus our comparison on matching the Doppler velocity of the maximum absorption observed during transit or at egress for the different solutions derived in the previous step.

Despite the large error bars, we found that the models with field strength 2 and 4 times the reference value give the best fits for the spectral position of the maximum absorption. That spectral position is defined by the average velocity of the projected bulk flow in the tail as derived in the previous step. For the B ~2.4 G case the velocity is ~54 km/s and for the B ~4.8 G case the velocity is ~49 km/s, which are the closest values to the maximum absorption spectral position observed for C II (~50 km/s). For the other strength values, the derived projected Doppler velocity (~ 22 km/s for the weak field case) is different to the value observed (Table 1), yet these cases cannot be fully rejected at this stage. We remark that we could obtain reasonably good fits to the C II line profiles for the case B ~2.4 G with a 30% tilted dipole, a degeneracy that cannot be actually removed because of the noise level of the data (e.g., Table 1).

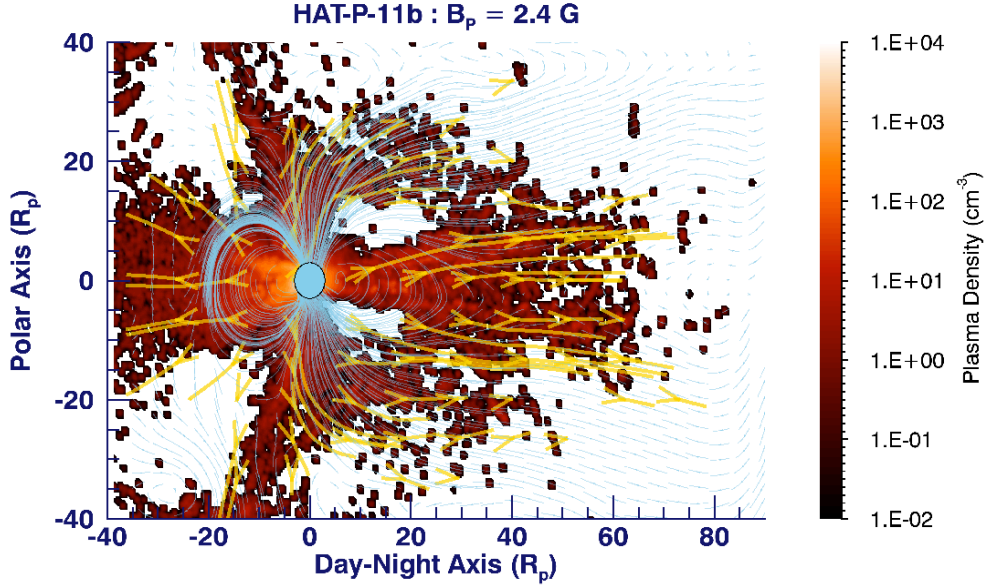

**Figure 5:** Flow streamlines (gold) over the density of C II ions that originally started their trip from the north and south polar cap regions. At high north and south distances, the polar wind outflow is bent tailward. Their presence of ions at lower latitudes is explained by the isotropy of the initial velocity distribution (Maxwellian) that injects a sub-population with velocity components perpendicular to the initial radial flow. That sub-population of ions are first trapped by closed field lines (blue) that are adjacent to the edge of the polar caps before they are quickly transported outward. Most ions shown, except the minor group flowing towards the star (see Supplementary Discussion III), contribute to the mass loading of the magnetotail by the polar wind.

**Step 4: C II and H I model consistency**

In this step we verify if the solution obtained above from the fitting of the C II transit absorption is consistent with the transit absorption observed for H I, particularly the Doppler position of the maximum absorption.

Our model is based on two H I populations: primary (no charge exchange) and secondary (single charge exchange) populations (e.g., Methods III.7). First, we find that the ~ 4% transit absorption due to the primary hydrodynamically injected H I population is insufficient to fit the observations for all atmospheric metallicities considered. However, after adding the secondary H I population (exactly same kinetic properties as the H II parents), we obtain a satisfactory fit to the Lya transit light curve (Fig. 2c; with the tail parameters indicated in Table 1) for *all* the relevant planetary magnetic field strengths retained from the C II analysis. To solve this degeneracy, we compared the model line profiles corresponding to the fits of the C II and H I light curves listed in Table 1 to the line profiles observed at specific orbital phases for which there is good signal-to-noise ratio S/N. Using a simple least-square fit over the spectral range affected by the transit absorption, we obtained a best fit for the planet magnetic field strength in the range 1.2-4.8 G for each of the three orbital phases corresponding to *HST* orbits 3 to 5. For the case B~0.1 G (1/10 times the reference value), we could not obtain a satisfactory fit to any of the three HST orbits (reduced $\chi^2$ in the range 7 to 20, Table 1), which rules out the weak-magnetization scenario for the planet. Since the ~9.6 G value is rejected by both the C II and H I analysis because their best fits require unacceptable size of the tail (e.g., Table 1), we are left with the range 1.2-4.8 G as the best solution for the magnetic field strength. Field strengths of ~1-5 G solution are consistent with the C II line profile analysis, although the diagnostic was not conclusive.

The strength of our approach resides in using physically motivated forward models that can reproduce the HST observations (see Methods, section IV), in including most known uncertainties related to both data and models that we carefully forecast in the overall errors attached to the solution for the B strength and the atmospheric metallicity (see Supplementary Discussion V ), and in proposing accurate predictions that future observations can test (see Methods, section IV & Supplementary Discussion VI ).

**Conclusions: a new paradigm to access bulk composition and magnetism of exoplanets**

With its low metallicity at most 3-times the metallicity of its star, and a magnetic field in the few Gauss regime, the properties of HAT-P-11b are akin to Jovian rather than Neptunian planetary properties, despite its small mass. Curiously the metallicity and magnetic field strength derived here are both in contradiction with previous planetary evolutionary modelling of HAT-P-11b that predicted a 56x solar (or 28x stellar) metallicity[23], and with scaling laws for B strength that do not account for the specificity of the planet[24] (e.g., Supplementary Discussion V).

An approach to solve the metallicity contradiction is to revise planetary evolutionary models by changing various ingredients (e.g., core mass, planet formation timing relative to the gas clearance of the disk, distance from the star, presence of nearby Jupiter-mass planet c, etc.) (e.g., Supplementary, Discussion VII). For the magnetic field, the situation is more complex than a specific scaling law for this population of "mini-Jupiter" exoplanets (e.g., Supplementary, Discussion VI), because the magnetic field strength is not easy to derive for any planet[24]. Such approaches could be interesting, yet they would miss the fundamental feature of the thousands of exoplanets detected to date: their diversity. If we keep the same paradigm for interpreting exoplanet metallicity and their magnetic field strength separately, we will end with as many scenarios and scaling laws as there are diverse exoplanets.

As far as a dynamo process is convection-driven (either thermal or compositional) in the deep interior of the planet (core or shell), an energy source is required, which could be related to a primordial heat (secular cooling from an initial hot state), or an ongoing differentiation due to settling of heavy materials toward the center[24,25], or even due to the magnetized parent star when orbiting close by. The altitude of the dynamo, the different layers that compose the interior of the planet, and the energy balance of the system are thus strongly tied to each other at every era of the evolution of the planet. We propose to expand the modelling shown here to the deep interior of a planet (down to the core), to include models of convection-driven dynamo processes, so that the conditions in the deep interior are consistent with every layer above, up to the magnetosphere. Our study shows the strength of the proposed comprehensive modelling. This new PanCET study has provided simultaneous constraints of two fundamental current properties of HAT-P-11b. Observations and modeling of other neutral and ionized species will be included in the future with probably more accurate constraints that formation and evolution models must fulfill (e.g., Methods, section IV, Supplementary Discussions VI & IX).

| B / B$_{ref}$ | 0.1 | 1 | 2 | 2T[a] | 4 | 8 |
|---|---|---|---|---|---|---|
| $|B_{eq}|$ (gauss) | 0.12 | 1.2 | 2.4 | 2.4 | 4.8 | 9.6 |
| $\theta$[b] (degree) | 40 | 26 | 21 | 21 | 18 | 13 |
| $\langle v \rangle_{tail}$ (C II) (km s$^{-1}$) | 22 | 27 | 54 | 57 | 49 | 27 |
| $\langle v \rangle_{tail}$ (H II) (km s$^{-1}$) | 49 | 93 | 112 | 86 | 111 | 72 |
| $\delta \langle v \rangle_{tail}$ (C II) (km s$^{-1}$) | 9.5 | 13 | 26 | 43 | 35 | 20 |
| $\delta \langle v \rangle_{tail}$ (H II) (km s$^{-1}$) | 15.5 | 48 | 62 | 55 | 78 | 87 |
| $\langle n \rangle_{tail}$ (C II) (cm$^{-3}$) | 4.7 | 0.3 | 0.33 | 0.39 | 0.15 | 0.06 |
| $\langle n \rangle_{tail}$ (H II) (cm$^{-3}$) | 8.4 10$^4$ | 1.3 10$^4$ | 9.3 10$^3$ | 6.3 10$^3$ | 7.2 10$^3$ | 0.3 10$^3$ |
| $\delta \langle n \rangle_{tail}$ (C II) (cm$^{-3}$) | 8.0 | 0.74 | 1.09 | 1.62 | 0.8 | 0.4 |
| *Magnetotail filling factor (C II)$^c$ (%)* | 97 | 48 | 56 | 34 | 27 | 7 |
| Metallicity (min, max) (times solar$^d$) | (1, 1.6) | (1, 1.6) | (2.0, 3.2) | (2.0, 2.8) | (2.5, 6) | (1, 3) |
| Mass loss rate[e] (hydro code; g s$^{-1}$) | 1.2 10$^{11}$ | 1.2 10$^{11}$ | 1.2 10$^{11}$ | 1.2 10$^{11}$ | 1.2 10$^{11}$ | 1.2 10$^{11}$ |
| Relative escape rate magnetosphere[f] (H & C) (PIC escape rate / Hydro escape rate; %) | 97 | 44 | 34 | 23 | 30 | 3 |
| Relative escape rate magnetosphere[g] (C) (PIC escape rate / Hydro escape rate; %) | 72 | 42 | 28 | 34 | 18 | 8 |
| C II tail tilt (degrees) | 0.1-2 | 0.1-0.4 | 0.1-0.3 | 0.1-0.3 | 0.1-0.4 | ≤0.1 |
| C II tail size (AU) | 0.4-0.8 | 1.8-3 | 2.5-9.3 | 2.5-7.5 | 1.9-3.1 | 6 - 19 |
| H I tail tilt (degrees) | 5-20 | 3-6 | 1-3 | 1-2.5 | 1-2.5 | ≤0.1 |
| H I tail size (AU) | 0.04 | 0.16 | 0.23 | 0.28 | 0.31 | 8.0 |
| $\chi^2$ orbit 5 (Lya line) | 14.8 | 1.3 | 1. | 1.0 | 0.75 | 0.7 |
| $\chi^2$ orbit 4[h] (Lya line) | 20.4 | 1.2 | 2.1 | 1.4 | 2.2 | 2.2 |
| $\chi^2$ orbit 3 (Lya line) | 7.2 | 0.6 | 1.6 | 0.7 | 0.5 | 0.9 |

**Table 1**: **Results of the sensitivity of the C II blue wing and H I Lya model transit absorptions versus the exoplanet magnetic field assumed, including the magnetotail size and tilt angle free parameters.** We highlight the case that is rejected because of the excessive size obtained for the magnetotail. Here B$_{ref}$=1.2 G. [a]Magnetic field axis with a 30° tilt from the spin axis of the exoplanet. [b]Polar cap cone semi-angle derived from PIC simulations (measured from magnetic field axis). [c]The volume filling factor is derived from the PIC simulation using the gas density distribution and taking into account the flattening of the magnetotail. [d]The metallicities are those implemented in the deep/low atmosphere 3D GCM and RT modelling and are constrained mainly

by the C II observations. Note that for formation and evolution scenarios, the star has two-times solar metallicity. **(e)**Mass loss rates correspond to the hydrodynamic escape from the upper atmosphere (hydro code). **(f)**We provide the ratio (%) of escape rate derived from the PIC simulations relative to the hydro code for the H and C species (H I, H II, C I, C II). It is remarkable to observe the decline of the mass loss rate versus the magnetic field strength. Most of the mass lost at the ionosphere boundary is recycled in the magnetosphere. **(g)**We provide the ratio (%) of escape rate derived from the PIC simulations relative to the hydro code for C II to check any differential escape for this heavy species. **(h)**Orbit 4 of the COS transit 2 shows slight distortion in various stellar line profiles, particularly in the Si III 120.65 nm line. This explains the large reduced $\chi^2$ values obtained for all B strength cases because the model line profile fit is based on the different stellar line profile of orbit 1 (e.g., extended data Figure 1).

## METHODS

Our novel approach (Fig.1) connects panchromatic observations of the HAT-P-11 system and a variety of interrelated models, linking the physical conditions in the planet's deep atmosphere (200 bar) to all layers above, including the magnetosphere and the stellar corona, to uncover the planet's metallicity and magnetic field strength. In section I, we focus on the global picture of our approach and particularly on its adequacy for the interpretation of the HST transit observations. In section II, we describe the HST observations and their calibration. In section III, we discuss each model in detail. In section IV, we assess the sensitivity of our results to model assumptions, emphasizing the impact of uncertainties on the final conclusions.

### I. Modeling and data analysis framework

Our framework starts with the 3D general-circulation modelling of the deep and lower atmosphere (pressure range 200-$10^{-5}$ bar), coupled with a radiative transfer scheme for different atmospheric metallicities[27]. Longitudinally averaged simulated composition, pressure-temperature (P-T) and eddy diffusion altitude profiles are then used as inputs to a more detailed 1D photochemical and thermochemical model of the deep, lower and middle atmosphere (P ~ 1000 to $10^{-9}$ bar), including both light and heavy species[28], and ion-chemistry. In the following step, the simulated species mixing ratios, P-T and eddy mixing profiles are injected into a 1D hydrodynamics and photo- and ion-chemistry model (hydro code) that describes the escape and transport of species in the upper atmosphere[29,30] up to few planetary radii.

To account for feedbacks between the middle and upper atmosphere, we iterate between models (Fig. 1) to achieve consistency between the different boundary conditions, particularly in the temperature profile (e.g., Methods, section III. 2 & 3 and Supplementary Discussion V). Assuming spherical symmetry, the ionized species of H II (protons) and C II derived from the hydro code are then injected at selected altitudes above $\sim 10^{-8}$ bar, where the atmosphere is no longer collisionally dominated, into the inner boundary of a 3D electromagnetic, relativistic and collisionless Particles-In-Cell (PIC) model of the magnetosphere[31,3,32] (Methods, section III. 4). We consider the day-night asymmetry of irradiation by reducing by a factor of two the flux of particles at the injection boundary radius on the night hemisphere (shadow region defined by the 1 planetary radius ($R_p$) "surface" of the planet).

In the outer boundary of the PIC model, the plasma and the interplanetary magnetic field (IMF) properties at the orbital position of the planet are derived with our 3D Magneto-Hydro-Dynamics (MHD) simulations of the stellar wind of HAT-P-11 (Methods, section III.5). In addition, we reconstructed the stellar radiation spectrum, from X-ray to infrared wavelengths, using both dedicated observations and modelling. This stellar spectrum is a key input to the atmospheric chemistry and dynamical models (Methods, section III.6).

Finally, we assume a Jovian-like dynamo process to produce a dipole magnetic field for the exoplanet described by its strength and tilt with respect to the spin axis. This is a realistic assumption for distances greater than 3-4 $R_P$, where the absorption signal is produced. We demonstrate this assumption has no impact on our conclusions, since any quadrupole or high order fields drop much faster from the body-center than the dipole field that effectively drive the magnetospheric structure (e.g., Supplementary Discussion VIII). With a radius $R_P \sim 2.78 \times 10^9$ cm, a spin rate of $\sim 1.488 \times 10^{-5}$ radians s$^{-1}$, and density of 1.33 g cm$^{-3}$, energetic considerations show that HAT-P-11b should sustain dynamo activity and possess a magnetic field even with an internal heat smaller than the internal heat of Neptune, yet the exact strength of the field is difficult to estimate[24] (Fig. 16 in Stevenson, 1983). Our simple dynamo model is described in Supplementary Discussion VI.

## II. HST data description, analysis, and calibration
### 1. Data description & analysis

We use data sets from two different HST programs: GO-14767 and t GO-14625. Each transit observation consists of five consecutive *HST* orbits covering exoplanet orbital phases before, during, and after the nominal transit time. Supplementary Table 1 lists the data log from the two HST programs. Four transit events were observed with the COS/G130M medium resolution grating (~113.0-146.0nm, which includes the C II 133.5 nm doublet) and two transits were observed with the STIS/G140M grating (~119.4-124.9 nm, including the H I Lya line). Although, the COS observations recorded the stellar Lya and O I 130.4 nm lines, these lines are fully contaminated by emissions from the Earth geocorona and airglow, respectively. All data were obtained in time-tag mode, which allows us to monitor any time variability related either to the instrument or stellar activity.

For the STIS/G140M data, we sub-sample each exposure in ~435 s sub-exposures using the recorded time-tag events. Each sub-exposure is then processed using the STIS calibration pipeline (CalSTIS) to obtain a 2D spectral image of the STIS long-slit (52´´x0.05´´), which is used to extract the stellar signal, after subtracting the Earth's geocoronal emission from adjacent areas of the detector along the spatial direction of the slit[33]. Most of the red wing of the stellar Lya line is absorbed by the interstellar medium (ISM) due to the large -63.24 km/s radial velocity of the star (Extended data Figure 2a), so only the blue wing of the Lya line is available for transit diagnostics.

Fortunately, the four *HST* visits made with COS sampled key far-UV stellar lines devoid of ISM absorption because the -63.24 km/s redshift of the star offsets any ISM absorption off narrow lines like the C II 133.5 nm doublet. There is also no Earth airglow contamination for C II 133.5 nm, that unfortunately contaminated the stellar O I 130.4 nm signal. To summarize, in contrast to Lya and the OI 130.4 nm triplet, the HAT-P-11 C II 133.5 nm doublet lines are not contaminated and are relatively strong, thus providing a complete and self-consistent diagnostic of the exoplanet's Doppler-shifted transit absorption of these ions over the full extent of the stellar line, in addition to directly monitoring the stellar chromospheric activity.

The C II doublet is actually an unresolved triplet composed of the 133.45 nm transition from the ground state (J=1/2) and two unresolved lines at ~133.57 nm that start from an exited state (J=3/2) and may be populated by collisional processes.

For the COS/G130M observations we used the default COS calibration pipeline, except for the statistical error estimation that we fully revise here (e.g., Methods, section II.2). Besides statistical noise, we also carefully checked that the transit absorptions are not related to stellar activity. We know that HAT-P-11 is an active, K2-K4V (effective temperature $T_{eff}$ ~ 4780 K), high-metallicity ([Fe/H] = 0.31) star[12]. Its chromospheric activity level defined by $R'_{HK}$ (the total flux in the Ca II H & K narrow bands normalized by the bolometric brightness of the star`), with log$R'_{HK}$ = -4.584, is comparable to the level log$R'_{HK}$ = -4.501 of the active HD189733 K0V star. Indeed, the chromospheric activity for HAT-P-11 recorded over a 450-day period with *Keck* shows constant high activity, with a ~ 10% modulation[10]. HAT-P-11 was in the *Kepler* field, and monitoring observations revealed that the exoplanet sits on a fairly polar orbit, evidencing a dynamically disturbed system[17,18,34]. The *Kepler* transit data revealed an active stellar disc with a latitudinal distribution of spots similar to sunspots (mean spots latitude ~16°), yet with a coverage area that is two orders of magnitude larger than for the Sun[35].

To track the stellar activity during the four *HST*/COS transit observations, we first use a diagnostic based on the C II emission lines. We compared the integrated flux of the C II 133.57 nm line to the C II 133.45 nm line versus the time (measured from mid-transit central time $T_c$; Extended Data Figure 3a) and versus the HST orbital phase. We find a nearly linear trend for both the C II 133.45 & 133.57 nm lines versus time that repeats for each of the five exoplanet orbital phases observed. For the out of transit observation (four data points around $T-T_c \sim -3.5$ hours, obtained between October 12 2016 and May 21 2017), the C II 133.57 and C II 133.45 nm emissions are strongly correlated with a Pearson coefficient ~ 0.91 and the integrated fluxes ratio $(I_{133.5}/I_{133.4})_{HAT-P-11}$~1.41 ± 0.05. For reference, for an optically thin atmosphere $(I_{133.5}/I_{133.4}) = 1.8$, which supports that for HAT-P-11, the two lines form in the moderately opaque region of the upper chromosphere and probably lower transition region[36]. This diagnostic is further confirmed by the shape of the two emission lines, both showing a single peak line profile with extended wings[36].

For the Sun, high spectral and spatial resolution observations[36] show C II lines that are also relatively optically thick emissions, forming between the upper chromosphere and lower transition region[36]. Using full-disk spectral-images (mosaics) gathered from the Interface Region Imaging Spectrograph (*IRIS*) scans obtained at distinct wavelengths between 2013 and 2017, we find that the disk-averaged C II 133.57 and C II 133.45 nm emissions are strongly correlated (Pearson coefficient of 0.999) over time, with a ratio $(I_{1335}/I_{1334})_{Sun}$ ~ 1.14 ± 0.02 that is comparable to the ratio derived here for HAT-P-11. Based on the strong correlation between the C II 133.45 and C II 133.57 nm emissions for both the Sun and HAT-P-11, we conclude that the C II 133.57 nm emission line can be safely used to monitor the variability of the stellar flux in that spectral range and correct for it at all orbital phases.

To further track the stellar activity during the transit observations of HAT-P-11b, we also generated light curves for strong far-UV lines (namely, Si III 120.6 nm and the average of Si IV 139.3 nm and SIV 140.2 nm), which are known to be good indicators of short-term variability due to the patchiness of the stellar disk for the Sun and for active stars like HD189733[24,26]. For both the Sun and HD189733, the C II lines show, by far, much less variability than the Si III and Si IV emissions (e.g., Figure 6 and Table 3 in Ben-Jaffel & Ballester, 2013[3]), a result consistent with the former lines being emitted by hotter layers of the upper chromosphere and lower transition region compared to the C II and OI lines[36].

For HAT-P-11, we first notice in the silicon lines a flare event that occurred during the fifth *HST* orbit of transit 2 (blue around T-$T_c$ ~ 2.75 hours) on December 16, 2016 (Extended Data Figure 3b). In contrast, the C II 133.57 nm line shows much less flare-related variability while the C II 133.45 nm line shows a different response to the flare, which supports that the variation observed for the C II 133.45 nm line, relative to the first out of transit spectrum, is not related to the stellar variability but to the exoplanet transit absorption as detected during the three other transits (red, black & green).

For instance, for transits 1 to 3, and except for the flare event discussed above, the Si III and Si IV light curves do not show any transit trend, which confirms that the light curve observed for the C II 133.45 nm line is not correlated with any stellar variability. The 2017 transit 4 (green) has a different behavior for Si III and Si IV emissions. Because it originates from a hotter layer (log(T) ~ 4.75) of the chromosphere, the Si IV lines are expected to show the largest variations, followed by the Si III line that originates from a cooler layer (log(T) ~ 4.25). In this frame, if the variation during transit 4 is of stellar origin, the scatter expected for the C II lines that originate from an even cooler layer (log(T)~4.1) should be smaller than observed for Si III lines *and follow the same temporal behavior*, which is not the trend observed during transit.

Our conclusion is that the stellar C II lines genuinely probe the transit absorption because we obtain the same repeated temporal trend over four distinct transit periods, despite the variations observed in other stellar lines. We considered the possibility of dismissing transit 4 and only use the first three transits in 2016 but found that this does not change our conclusions apart from slightly increasing the statistical error bars. All these pieces of evidence strongly support the detection of the transit absorption for the blue wing of the C II 133.45 nm line.

It is interesting to note that the H I and C II absorption features overlap on spectral ranges where enough signal is available for both H I Lya and C II 133.45 nm lines. For example, the spectral window -150 km/s to -30 km/s used for the Ly-a detection overlaps with the C II 133.45nm line over
the -70 to -30 km/s range, where enough signal is available (up to ~$2.5 \times 10^{-14}$ ergs cm$^{-2}$ s$^{-1}$ A$^{-1}$) in the CII line. Similarly, the spectral window -70 to -10 Km/s used for the C II detection also overlaps with the window available for H I. As shown in Figure 2A, around Doppler position -50 km/s, the Lya flux ~ $3 \times 10^{-14}$ ergs cm$^{-2}$ s$^{-1}$ A$^{-1}$ (close to the line's peak signal) is large enough for any potential absorption feature. We conclude that the observed velocity difference is not caused by the fact that we cannot observe the same velocity range in both lines.

## 2. COS calibration pipeline: Statistical noise reassessment

*HST* COS was designed to work in the very low count rate regime, using Poisson statistics to evaluate error bars $\sqrt{N}$ that are attached to the measured counts $N$. This convention was applied to all archived COS data up to the end of 2012. The COS calibration pipeline (CalCOS 2.19.1 and later) implemented since 2013 used a new prescription to estimate statistical errors based on Gehrels (1986)[37], intended to correct for the limiting case when the signal counts and corresponding error are close to zero. With the approximate upper bound of the confidence interval defined by[37]

$$L_u = N + 1 + \sqrt{(N + 3/4)} \quad \text{Eq. (1)}$$

its error is estimated as $Err^{(u)} = 1 + \sqrt{(N+3/4)}$. This recipe produces an asymmetric confidence interval with respect to the mean value, yet nothing was stated in the *HST*/COS handbook on the lower boundary of the confidence interval that was initially estimated:

$$L_b = N - \sqrt{N} \quad \text{Eq. (2)}$$

with an error $Err^{(b)} = \sqrt{N}$ that is no longer used in CalCOS.
For reference, in the current CalCOS version, there is no straightforward way to implement any other expression of the error bars except for the default one. To illustrate the strong effect of the currently implemented error expression (Eq. 1), we show in Extended Data Figure 4 errors obtained before and after 2013 for the HD 209458 spectrum obtained with the COS/G130M grating[38]. If one keeps the inflated statistical noise actually estimated in the current CalCOS pipeline, we miss many detections for faint targets.

To remedy the problem of low or zero counts detection, Gehrels assumed single-sided upper and lower confidence limits and used the relation between the Poisson and $\chi^2$ probability functions to derive simplified approximations for each bound of the confidence interval (e.g., Eq. 1 & 2). One may also consider double-sided confidence intervals but with a lower confidence level than when considering each side of the interval[37]. The problem is that for low counts signals, the size of the proposed error is comparable to the signal itself, which leads to very low S/N ratio across the entire spectrum that becomes difficult to detect.

Estimating confidence intervals for a Poisson mean is an old problem in statistics. Standard exact confidence intervals tend to be very conservative and too wide, particularly for moderate count levels[39]. Our approach here is to favor methods that make the confidence interval the narrowest[39]. Because we are interested in using the confidence interval as a measure of the statistical error, we propose to use the two-sided confidence limits instead of the one-sided confidence limit developed in Gehrels (1986)[37]. In this framework, a few interesting solutions appear when using the classical 68% confidence level, such as the so-called two-sides Rao score confidence limits[39]:

$$L_{u/b} = N + 1/2 \pm \sqrt{(N + 1/4)}, \quad \text{Eq. (3)}$$

with an upper bound error $Err = 1/2+\sqrt{(N + 1/4)}$,
or the expression proposed in Barker (2002)[40]

$$L_{u/b} = N + 1/4 \pm \sqrt{(N + 3/8)}, \quad \text{Eq. (4)}$$

with an upper bound error $Err = 1/4+\sqrt{(N +3/8)}$,
or the so-called continuity corrected Wald interval[40]:

$$L_{u/b} = N \pm \sqrt{(N + 1/2)}, \quad \text{Eq. (5)}$$

with an upper bound error $Err =\sqrt{(N +1/2)}$, where $\pm$ are for the upper/lower edges of the confidence interval.
 Interestingly, the new errors (Eq. 3-5) fulfill the same constraints—namely, they tend to the standard deviation for large $N$ counts and give a finite value at $N=0$. However, the confidence interval is now much narrower than the one used in CalCOS.
 In the future, we recommend implementing Eq. 5 in the CalCOS pipeline, or simply going back to the classical standard deviation.

### III. Comprehensive global modelling of exoplanetary atmospheres

Next, we provide a detailed description of our framework for the study of the HAT-P-11 system. It includes all levels of the atmosphere and external environment of the exoplanet, starting from the deep interior and reaching up to the magnetosphere and stellar corona (see Fig. 1). Below, we describe the models separately developed for each layer and how they connect through boundary conditions.

#### 1. Deep-lower atmosphere: 3D GCM code to simulate the thermal structure and eddy diffusion profile

To model the thermal structure and vertical mixing in the deep-lower atmosphere of HAT-P-11b (P=200 bar -0.01 mbar), we utilize the Stellar and Planetary Atmospheric Radiation and Circulation (SPARC) model, which couples a two-stream, non-gray radiative transfer code by Marley and McKay (1999)[41] with the 3D General Circulation Model (GCM) MITgcm[42]. MITgcm employs the primitive equations, a simplification of the fully compressible fluid equations assuming hydrostatic balance. The radiative transfer code is used to solve for the upward and downward fluxes at each grid point, which in turn are used to derive heating rates to update the wind and temperature fields in the dynamics. SPARC has been extensively used to model the atmospheric circulation of hot Jupiters[27,43], sub-Neptunes[44], and super- Earths[45].

For each simulation of HAT-P-11b, we utilize a cubed-sphere grid with a horizontal resolution of C32 (approximately equivalent to 64x128 elements in latitude and longitude) and 53 vertical levels. We model six atmospheric compositions: 1, 3, 5, 10, 30, and 50 times solar abundances. For each metallicity case, all species aside from $H_2$/He are enhanced by their respective factors. Opacities are calculated at each temperature/pressure point assuming local chemical equilibrium and accounting for condensates rainout, using Lodders (2003)[46] elemental abundances.

We estimate eddy diffusion coefficient ($K_{zz}$) profiles in the middle atmosphere using the root-mean-square (*rms*) vertical velocities derived from the GCM simulations of HAT-P-11b, by calculating $K_{zz} = w(z)L(z)$. Here, w(z) is the limb-averaged *rms* vertical velocity and L(z) is the atmospheric pressure scale height, both as a function of altitude.

To save time, we also used the 1D atmosphere ATMO model[47] to generate a forward atmospheric model for the lower atmosphere (100 to $10^{-5}$ bar). ATMO computes the 1D temperature-pressure (T-P) structure of an atmosphere in plane-parallel geometry in radiative, convective, and chemical equilibrium. ATMO includes isotropic multi-gas Rayleigh scattering and $H_2$-$H_2$ and $H_2$-He collision-induced absorption, as well as opacities for all major chemical species taken from the most up-to-date high-temperature sources, including $H_2O$, $CO_2$, CO, $CH_4$, $NH_3$, Na, K, Li, Rb and Cs, TiO, VO, and FeH. We generated T-P profiles using 32 correlated k-bands across the 0.2 $\mu$m to 1 cm wavelength range, evenly spaced in wavenumber. Spectra were generated using 5000 correlated k-bands across the same range to resolve spectral features. We used a stellar model for input flux from the host star (e.g., Methods, section III.6). Rainout chemistry was treated following[46,48] Burrows & Sharp (1999), with precipitation depleting condensable species at pressures where the T-P profile crossed the corresponding condensation curve and also at higher altitudes. We set the heat redistribution factor to $f = 0.5$, which assumes complete redistribution, calculating models for 1 x and 50 x solar abundances.

We use both thermal and eddy diffusion altitude profiles as input to the following modeling steps of the middle and upper atmospheres.

## 2. Lower-middle atmosphere: 1D kinetic photo- and thermo-chemical model

We calculate the chemical composition of HAT-P-11 using a 1D photochemical-thermochemical model[28]. The model solves kinetically for thermochemical equilibrium in the whole atmosphere, taking into account atmospheric mixing, molecular diffusion, and stellar radiation. The thermochemical equilibrium that dominates in the deep atmosphere is described through the microscopic balance of multiple chemical reactions that include species of H/C/N/O/S composition. At lower pressures (different for the various species, but roughly P < 10 - 1 bar), the equilibrium is perturbed through atmospheric mixing (described through an eddy mixing profile) and photochemistry. Our modeling includes ion chemistry and spans the P= 200 – 1 nanobar regime.

The model requires as input a thermal vertical structure profile for which we combine results from GCM models for the lower atmosphere averaged over the whole planet, with results from the 1D atmospheric escape simulations for the upper atmosphere (Extended Data Figure 5). We smoothly join profiles from the lower and upper atmospheres, taking into consideration atmospheric stability—i.e., we verify that the atmospheric lapse rate for the assumed temperature structure is sub-adiabatic everywhere. For the assumed atmospheric mixing, we use input for the simulated $K_{zz}$ profiles derived from the GCM. For the upper atmosphere, we consider a constant $K_{zz}$ profile at the value defined from the GCM results. From the perspective of the upper atmosphere, the altitude profile of $K_{zz}$ is not critical as long as the resulting eddy mixing is large enough as to suppress the heterosphere at the higher altitudes where it might otherwise be formed. It is likely that there is no homopause in the atmosphere of HAT-P-11b and that separation by mass does not occur. Eddy mixing is superseded by vertical advection at higher altitudes, leaving no role for molecular diffusion. In other words, profiles of $K_{zz} \geq 10^7$ cm$^{-2}$ s$^{-1}$ will produce the same fluxes of heavy elements in the upper atmosphere. At pressures higher than 100 bar, we assume that the eddy efficiency will increase due to convection. However, simulations with monotonic eddy profiles in the lower atmosphere demonstrate that the $K_{zz}$ values below 100 bars do not modify our compositional results. Both thermal structure and eddy mixing will change depending on the assumed elemental composition. For the different metallicity cases we study, we use profiles that are interpolated from limiting cases of 1x and 50x solar metallicity. Extended Data Figure 5a, b shows the vertical distributions of typical species in the lower/middle atmosphere of HAT-P-11b for the 1x solar metallicity reference case.

## 3. Upper atmosphere aeronomy: 1D photochemistry and hydrodynamic transport code (hydro code)

The aeronomy model is described in García Muñoz (2007)[29,30]. It solves for mass, momentum, and energy conservation in the planet's thermosphere-exosphere. The formulation assumes that the hydrodynamic outflow (resulting from stellar XUV irradiation) is spherically symmetric. The atmosphere is irradiated at zero zenith angle. To account for partial shadowing of the nightside, we reduced the outflow density by a factor of two (e.g., Supplementary Discussion V) We also tested other irradiation zenith angles corresponding to limb-average conditions relevant to transit observations. The model incorporates both neutral and ion photochemistry.

The bottom and top boundaries are placed at a pressure of 10 μbar and at ~18 planetary radii. The simulations were carried out with the H-He-C-O-N-D-CH chemical network[29], which includes 46 species of hydrogen, helium, carbon, oxygen, nitrogen, and deuterium in 223 chemical reactions. The chemicals are split into 19 neutral species and 27 charged species, including molecules, atoms, and thermal electrons. The Lagrangian $L_1$ point that separates the domains where the gravitational field is dominated by the planet or the star is at about ~7 planetary radii above HAT-P-11b's optical radius. The chemical species in the model are transported by bulk-gas advection, eddy, molecular, and ambipolar diffusion[28].

The model[29] was upgraded to account for cooling by H I atoms excited in electron collisions[49], and a newer formulation of $H_3^+$ infrared cooling[50]. Both H I and $H_3^+$ potentially behave as thermostats at high temperatures. However, their impact on the HAT-P-11b simulations is relatively minor.
The implemented eddy diffusion coefficient ($K_{zz}=2\times10^{10}$ cm$^2$ s$^{-1}$, independent of altitude) is based on the mixing efficiencies inferred from the GCM at overlapping pressure levels. Strong eddy mixing prevents the occurrence of a heterosphere on HAT-P-11b. Eddy diffusion and advection are the dominating transport mechanisms for bulk gas densities larger and smaller than about $4\times10^9$ cm$^{-3}$, respectively. At the bottom boundary, the aeronomy model adopts the gas concentrations calculated by the lower-atmosphere photochemical model at 10 μbar (e.g., Extended Data Figure 5.

Supplementary Table 2 lists the adopted volume mixing ratios at the bottom boundary, together with the calculated mass loss rates for all the solar metallicity conditions considered (1, 2, 6, 10, 30, 50x solar). Extended Data Figure 5c shows the vertical distributions for typical species in the upper atmosphere of HAT-P-11 b for the 1x solar metallicity reference case. The derived temperature profile in the upper atmosphere is not too sensitive to the assumed metallicity (Extended Data Figure 5a). For all metallicities, the bulk flow quickly becomes supersonic at distances above ~2 Rp (Extended Data Figure 5d).

### 4. Plasmasphere & magnetosphere: PIC E-M 3D code to simulate the plasma distribution around the exoplanet

Here, we use a PIC electro-magnetic/relativistic 3D code, built and validated for the Earth and Mercury magnetospheres[31,51], Earth polarwind (Barakat & Schunk, 2006)[52], and recently extended to hot Jupiters HD189733b, WASP-12b, and potential exomoon tori[3,32].

Electrons and ions are represented as macro-particles, each containing a large number of real particles. The code solves the Maxwell Equations on a 3D grid:

$$\frac{\partial \boldsymbol{B}}{\partial t} = -\Delta \times \boldsymbol{E}$$
$$\epsilon_0 \frac{\partial \boldsymbol{E}}{\partial t} = \mu_0^{-1} \nabla \times \boldsymbol{B} - J$$

where $\boldsymbol{J}$ is the current vector, and follows each macro-particle in the simulation box using the Newton-Lorentz motion equation:

$$\frac{d(\gamma m \mathbf{v})}{dt} = q(\boldsymbol{E} + \mathbf{v} \times \boldsymbol{B}) + \boldsymbol{F_G}$$

where $\boldsymbol{F_G}$ is the gravity force and $\gamma = \sqrt{(1-(v/c)|2)}$ is the relativistic motion factor.

The technical difficulties inherent to the huge contrast between plasma kinetic scales and the macroscopic scales of the magnetosphere have been extensively discussed in the literature[3,32,33,51,53,54].

The way to address the problem was to scale the plasma parameters in order to shrink the computing time while keeping most of the physics needed for the macro-system. The adequacy of using a PIC code to study a magnetosphere is discussed in details in Supplementary Discussion X.

To answer the question of whether the kinetic spatial scales of the ions are resolved in the PIC model, we derived the gyroradius (defined as $m_i*v_\perp/(q_i*B)$, where $m_i$ is the ion mass, $v_\perp$ is the local ion's speed perpendicular to the field, $q_i$ its charge, and B is the local magnetic field), of the two main ions (C II and H

II) considered in our study in the XZ plane (Extended data figure 6). The plot shows that the gyroradius of each species is well resolved and the macro-ions have enough space in the simulation box to interact with the magnetic field and complete their gyration motion. In the magnetotail, the gyroradius of macro-ions becomes larger (weaker magnetic field), consistent with the general picture that species are escaping along the tail on straight trajectories. For the stellar wind plasma, the macro-protons enter the simulation box with a gyroradius as large as ~30 $\Delta$, which shrinks to small values as soon as the particles start feeling the dipole field. For all plasma sources, the gyroradius for macro-electrons is even smaller by the mass ratio $m_i/m_e=100$. Therefore, all those scales are well resolved in our simulation, properly describing charge separation and kinetic acceleration of species.

Radiation pressure forces are likely important for the distribution of H I atoms with a moderate opacity but may be safely neglected for C II ions because the C II stellar line is much fainter than the stellar H I Lya line, and the C II ions are also heavier. For reference, the maximum value of the ratio of radiation pressure force to gravity force for HAT-P-11 is $\beta_{max}$ ~ 0.02 for C II[55], which is too small to affect the dynamics of C II ions that are governed instead by strong E-M forces and the complex magnetospheric current system.

In the PIC code, we adopt the exoplanet magnetic field strength with the range of values assumed in our sensitivity study (see Main Article). Stellar wind properties at the orbital position of the planet are derived from MHD 3D models of the stellar wind (Methods, Section III.5). The exoplanet's (optionally tilted) magnetic field is assumed to be dipolar. The ion-to-electron mass ratio $m_i/m_e=100$ for macro-protons is large enough to obtain a good separation between opposite charges[56]. The code parameters are selected to yield an ion skin depth that ensures the magnetospheric cavity is properly resolved with the selected grid ($\Delta r$ = 0.33 $R_p$), where r is distance from the planet center[57,58]. The grid fulfills the *Courant* condition $c\Delta t < \Delta/\sqrt{3}$ (c=0.5 is the speed of light and $\Delta t$ is the step time in the code), which helps avoid numerical instabilities[3,59]. We also adopt a strong condition on the plasma frequency $\omega_p \Delta t < 0.25$, which efficiently reduces plasma instabilities[3,59]. In addition, we avoid the problem of grid heating[3,59] by enforcing that the Debey length remains larger than a critical level $\lambda_D \geq \Delta r/\pi$.

To obtain shielding of charges over the Debey volume, we load five pairs of particles per simulation cell[3,59]. Macro-ion and macro-electron pairs are randomly and continuously injected to reproduce a spherical outflow around the exoplanet (planetary wind) with the radial kinetic temperature, density, and bulk speed provided by the hydro code (e.g., Methods, section III.3). From the moments of the macro-particle velocity distributions, we derive the plasma number density, temperature, and bulk velocity. To calculate the transit absorption, the box distribution is re-oriented in 3D to take into-account the aberration angle of the nose-magnetotail orientation or any small tilt in the planet's magnetic field ($B_p$) with respect to the spin axis of the rotating planet.

For the HAT-P-11 parameters (e.g., Methods, section III.5), the stellar wind is super-magnetosonic at the orbital position of the exoplanet. We use a 3D cartesian simulation box centered at the planet's position in the OX-OY-OZ directions, where OX is the star-planet line, OY is the dawn-dusk direction, OZ is the spin axis.

The stellar wind particles impinge on the OYZ plane, resulting in a total of ~$1.8 \times 10^8$ macro-ion and macro-electron pairs in the box. To produce the planetary wind, we inject planetary macro-protons ($m_i/m_e$ =100) and ionized macro-carbon ($m_{Cii}/m_p$=12) paired with their corresponding macro-electrons, using the species altitude profiles derived from the hydro code. Initial conditions require a Maxwell distribution for all species at the temperature provided by the hydro code. In total, a maximum of ~$8 \times 10^7$ macro-proton and macro-electron pairs, along with a maximum of $8 \times 10^7$ macro-C II and macro-electron pairs, are injected in the system. When required, the code separately tracks a sub-population of any family of particles (like following C II and their electrons that escape from the exoplanet's polar caps). In the general case, the pressure level (generally above ~$10^{-8}$ bar) of the bottom boundary layer of the PIC simulation depends on each atmospheric model used and is derived using the altitude level where electromagnetic forces take over collisional forces, leading to the decoupling between ions and neutrals[60]. A simple diagnostic to check the location of that boundary is to estimate the altitude position (~1.1 to 1.4 $R_p$) where the ionization fraction is sufficiently high (electron volume mixing ratio $x_e > \sim 10^{-3}$)[60]. Finally, fields are able to propagate into space without reflection on the facets of the simulation box[61,62].

We use our PIC code results to investigate planetary magnetic field lines and typical plasma distributions of different origins for one of our best-fitting $B_p$ ~2.4 G (Figures 4 & 5). For the stellar wind plasma (impinging from the left of the figure), we derive the classical structure with a standoff distance at the magnetopause nose located at ~12 $R_P$ (dayside, Fig. 5c), but with an extended magnetotail (nightside) where a reconnection appears in the field lines around ~50 $R_p$ tailward (Fig. 5d). The final configuration is an open magnetosphere with extended parallel field lines on the night side. We also note the precipitation along field lines of stellar wind particles from the magnetotail equatorial sheet back to the planetary poles (Fig. 5c). For the planetary source, our code recovers the corotation dynamics expected in the plasmasphere and the resulting strong tailward outflow (Fig. 5a, b)[26]. Finally, our PIC code reproduces many features like the polar outflows (the exoplanet polar wind Fig. 5), and the cross-field planetary wind predicted since the 1990s[63] (Figures 4a, b & 5).

## 5. Stellar wind plasma & Interplanetary Magnetic Field conditions at the orbital position of the exoplanet: MHD 3D code simulations

The stellar wind conditions at the planetary orbit are calculated using the Alfvén Wave Solar Model[64]. The model calculates the non-ideal MHD solution for the stellar corona and stellar wind, taking into account coronal heating and wind acceleration by Alfvèn waves, as well as coronal thermodynamics, radiative cooling and electron heat conduction. The model is driven by observations of the surface radial magnetic field of the magnetograms for the solar case, and Zeeman-Doppler-Imaging[65] (ZDI) for the stellar case. This approach has been used in stellar corona and wind simulations of various systems—e.g., Sun[66], HD189733[67], etc. No magnetic field data are currently available for HAT-P-11. We searched for the most similar system with available ZDI data and found HD189733, with the following parameters for HD189733 (former) and HAT-P-11 (latter): Spectral type=K1.5, K4, Age=~1, 6+6/-4 Gyr, $R_{star}$=0.8, 0.7 $R_{sun}$, $M_{star}$=0.8, 0.7 $M_{sun}$, P=12, 29 days. Recent work on HD189733 confirms an average equatorial stellar magnetic field strength of ~30 G[68]. However, HAT-P-11 might be older, and it is a slower rotator although the stellar Ca II chromospheric activity is lower than HD189733's, but not as low as the average Sun. Using standard correlations between surface magnetic field strength and age, and strength and rotation[69], we derive a field strength of ~1-2 G for HAT-P-11, similar to the Sun. We considered both a scaled ZDI magnetic field map of HD189733, and the magnetic field map for solar maximum activity conditions [Carrington Rotation map (CR1962) around year 2000]. We ran our MHD model for both magnetic field conditions but with the intrinsic parameters of HAT-P-11. At the exoplanet orbit around the transit's phase, the two MHD simulations converge on stellar wind parameters in the range $(1.3-1.5) \times 10^6$ K for the temperature, 500-600 km/s for the speed, and $~3.3 \times 10^3$ cm$^{-3}$ for the density (e.g., Extended Data Figure 7 for the temperature distribution).

The MHD 3D-predicted coronal temperature is $~3 \times 10^6$ K, which compares with the temperature range derived from *XMM* X-rays observations of HAT-P-11 (see below). We also tested the Parker model with coronal temperatures derived from the X-ray observations and found stellar wind speeds at the exoplanet orbital position similar to those from MHD simulations. The MHD 3D code is, however, superior because it provides the stellar wind (SW) plasma and field variable conditions at the exoplanet's position along its eccentric and nearly polar orbit.

We note that without direct stellar wind data, and with the uncertainties associated with the transit observations (e.g., Methods, sections II), and the large number of parameters describing the star-planet system, a detailed study of the stellar corona and wind of HAT-P-11 becomes impractical. We thus focus on a range of stellar parameters that are consistent both with X-ray observations and MHD 3D models, leading to the stellar wind ram pressure and plasma properties in the range displayed in Supplementary Table 3.

### 6. Stellar radiation inputs:
#### 6.1 Stellar XUV spectrum reconstruction

The absorption of photons with λ < 912 Å in the interstellar medium (ISM) prevents the detection of extreme-UV radiation (λ ~100 - 912 Å) from almost every star. The XUV (~1 - 912 Å, X-ray + extreme-UV) spectrum of a late-type star is dominated by continuum and emission lines originating from the material at log T (K) ~ 4-8 present in the transition region and corona[70]. To model the spectral energy distribution (SED) in the XUV, we built a model of the emitting material in these layers, using X-ray spectra originated at the hottest temperatures and far-UV spectral lines formed at lower temperatures[70].

We use *XMM-Newton* data, complemented with the *HST* far-UV spectrum from this work. *XMM-Newton* observed HAT-P-11 in May 19 2015 (Obs. ID 764100701), using the three EPIC cameras (EPIC-pn, 16.9 ks, EPIC-MOS1 28.9 ks, EPIC-MOS2 29.1 ks), with a combined S/N ratio of 13.2. We fit spectra following standard procedures within ISIS[71], the Interactive Spectral Interpretation System software and complemented the coronal (and transition region) model with the far-UV line fluxes measured.

We generated a synthetic SED in the range λλ 1 – 1200 Å using this model (e.g., Supplemental Discussion XI for more details). The derived extreme-UV (10-92 nm) luminosity is $L_{extreme-UV}$ (erg s$^{-1}$) ~ 4.01 x 10$^{28}$ erg/s, and the X-ray part (0.5-10 nm) is Lx ~ 2.36 x 10$^{27}$ erg/s, which are consistent with early calculations[72]. The final XUV spectrum is shown in Extended Data Figure 2b.

#### 6.2 Stellar FUV-IR spectrum reconstruction

HAT-P-11 is a K2-K4 V main sequence star. To construct its full spectral radiation flux, we start from the stellar spectrum of Eps Eri (K2 V) that was reconstructed from observations and provided in the MUSCLES database[73]. As a first step, we subtract a PHOENIX continuum model (BT-NEXTGEN, 2009; $T_{eff}$ = 5000, Log(g) = 4.5, solar metallicity) that fits the Eps Eri spectrum in the long wavelength range, and add another PHOENIX model with the HAT-P-11 parameters ($T_{eff}$ = 4700, Log(g) = 4.5, two times solar metallicity). Because we have no near-UV observations for HAT-P-11, nor observations of the longest wavelengths in the far-UV, we keep the same observed Eps Eri flux for that spectral range (143.0-300 nm) after correcting for the distance and size of stellar discs (see Supplementary Discussion XI for more details).

For the far-UV range (115.0-143.0 nm), we use our COS G130M observations for most lines and STIS 140M observations to reconstruct the Lya line (Extended Data Fig. 2a). Because of the large radial velocity of the star, most thin lines, like in the C II 133.5 nm doublet, are not affected by the ISM absorption. However, the Lya line is still strongly affected because it is particularly broad. Starting with a symmetric intrinsic stellar Lya line[74], we obtain a reasonable fit to the observed line profile with an ISM H I column density of ~ $5\times10^{18}$ cm$^{-2}$ along the line of sight toward the star (Extended Data Figure 2a). For the XUV range, we replaced the Eps Eri spectrum with the one reconstructed in Methods, section III.6.1. We used the final full stellar spectrum, shown in Extended Data Figure 2b, as input for all the theoretical modeling used in the present study.

7. **Lya transit interpretation and modeling: Approximation of two H I populations**

The H I Lya transit has been used extensively as a direct diagnostic of the evaporation and related mass loss from exoplanets[1,2,4]. Several scenarios for explaining the observations have been proposed, mainly related to thermal atoms absorption, ENAs production or radiation-accelerated atoms[11,33,75,76,77].

Here, we compare the absorption by a spherical cloud derived based on the H I radial distribution obtained by our hydro code, to the ~ (14-32) % Lya transit depth observed by *HST*/STIS for HAT-P-11b (e.g., Fig. 2). Our results show that for 1x-150x solar metallicity, the model Lya transit absorption, ~ (4.2-3.9) %, falls short of the observed level for the spectral range shown in Figure 2.

To take into account the key properties of the plasma distribution around the exoplanet, we consider a simplified model with only two populations of neutral atomic hydrogen. The first is the primary H I population from the hydro code described above. The secondary H I population is created by the first resonance charge-exchange reaction between one proton of the plasma in the magnetosphere (of planetary or SW origin) and one neutral of the primary H I population[78]. In this process, the newly created neutral H I has the same velocity distribution as the parent proton (*p$^+$ + H I → *H I + p$^+$). Ripken and Fahr (1983)[79] approximate the production rate of neutrals from charge exchange as:

$$P(r,v) \sim \sigma(v_{rel}^+) * v_{rel}^+(r, v) * n_{H\,I}(r) * f_{p+}(r, v) \qquad \textbf{Eq. (6)}$$

where $v_{rel}^+(r, v)$ is the average relative velocity of all neutrals with respect to protons of velocity $v$, $n_{H\,I}(r)$ is the density of primary H I (from hydro code), $f_{p+}(r, v)$ is the velocity distribution of protons (PIC code), and $\sigma(v_{rel}^+)$ is the charge exchange cross section which is velocity dependent[80] (Maher and Tinslay, 1977). For the destruction rate we use a similar approximation but for all protons with respect to neutral atoms of velocity $v$:

$$L(r,v) \sim \sigma(v_{rel}^-) * v_{rel}^-(r, v) * n_{p+}(r) * f_{H\,I}(r, v) = \Gamma_{ext}(r, v) * f_{H\,I}(r, v) \qquad \textbf{Eq. (7)}$$

where $\Gamma_{ext}(r, v)$ is the destruction frequency[78]. The production term that contributes to the transport equation for the secondary population (Eq. 6) is proportional to the H II ion velocity distribution, whereas the destruction term (Eq. 7) is proportional to the number density of the same ions.

Another limitation of the model used here is the neglect of radiation pressure and stellar gravity[75], an acceptable approximation when the hydrogen cloud is optically thick as it is the case here[76]. Our simplified model for the Lyα transit absorption demonstrates the importance of taking into account magnetospheric processes, yet more work is needed to obtain a self-consistent model of the H I distribution around a magnetized outflowing exoplanet. A full kinetic treatment of the problem in 3D, coupled with the 3D PIC plasma code is underway and will be presented in a forthcoming study.

### IV. Sensitivity to model assumptions & overall error, and the robustness of the results

In Supplementary Discussion V, we address the importance of feedback between modules (described in Methods, section III) and how taking them into account impacts our conclusions. In addition, we assess the sensitivity of our results to model assumptions and evaluate the corresponding impact on the overall error. The arguments provided in Supplementary Discussion V reinforce the robustness of our results regarding the B strength of HAT-P-11b and its atmospheric metallicity.

Here, we summarize the main strengths of our approach:
- Our finding about the low metallicity of HAT-P-11b is consistent with results from an independent and simultaneous study using spectrally-extended optical/IR HST transit observations and a quite different approach that probes the planet's lower atmosphere[13].
- Our global model predicts the right Doppler-shift speed for two distinct species (CII & HI), particularly the speed ratio (~ 2), which cannot be explained by simple considerations. In addition, this ratio seems consistent with similar finding reported for OII and HII speeds with a ratio ~2.6 that is observed in the polar wind at ~9 $R_{Earth}$ from Earth[81].
-  We provide enough details and predictions that can be further tested with future observations/modeling. For example, the phase-extended light-curve and the spectral shape of the transit absorption for distinct species can be immediately improved with dedicated HST observations. Also, the expected internal energy that we predict to sustain the dynamo process of the planet can be tested with future JWST IR observations of the exoplanet thermal emission during secondary eclipse (see Supplementary Discussion VI). Finally, we provide enough details about the plasma properties in the HAT-P-11b atmosphere-magnetosphere that can be cross-checked with distinct simulation tools like using multi-fluid MHD or hybrid codes.

All arguments discussed above reinforce the robustness of our results and give us enough confidence in the reported conclusions.

**Data Availability.** The data that support the plots within these paper and other findings of this study are available from the corresponding author upon reasonable request. HST reduced data are available to the public through archives.stsci.edu using the dataset names shown in Supplementary Table 1. An ascii version of the HAT-P-11 stellar spectrum shown in Extended Data Figure 2b can be downloaded here https://doi.org/10.48392/lbj-001

**Code Availability.** All the codes used in this study have been employed in the past for published work and references are provided in the manuscript. Those references include enough detail as to make the model predictions reproducible. The PIC code is an old version of the Tristan code that is available to public though github: https://github.com/ntoles/tristan-mp-pitp.

**Acknowledgements**

We are very grateful to the staff at the Space Telescope Science Institute (STScI), in particular to Patricia Royle, Steven Penton, William Januszewski and John Debes, for their careful work in the scheduling of our observations and for various instrument insights. We also thank Carlos Carvalho at IAP and Terry Forrester at LPL for the set-up and managing of various computing resources for this project. The authors thanks V. Bourrier, David Ehenreich, and A. Lecavelier for helpful discussions about the analysis of the Lyα transit data. L. Ben-Jaffel and P. Lavvas acknowledge support from CNES (France) and CNRS under project PACES. All US co-authors acknowledge support from STScI under the PanCET GO 14767 program, and G. Ballester, G. Henry, and T. Kataria were also funded under GO 14625.

This work is based on observations with the NASA/ESA *HST*, obtained at the Space Telescope Science Institute (STScI) operated by AURA, Inc. This work is also based on observations with *XMM-Newton*.

**Author Contributions:** L.B.J. led the data analysis with contribution from G.E.B, and identified the problem and solution for the current COS pipeline derivation of errors. He defined and planned over time the 3D PIC magnetospheric simulations and the post-processing of the large set of plasma data that enabled the reported analysis. He also defined the H I interactions with protons. He defined the final strategy and reasoning to disentangle the different parameters of the problem and get the different results summarized in Table 1. He used a Parker model for the stellar wind definition. He also derived a simplified model regarding the magnetic field strength and internal heat of the planet. He prepared the final draft with contributions from G.E.B. in particular and other co-authors as described below.

G.E.B. initiated the multifaceted far-UV approach to study of HAT-P-11b in heavy metals versus H I, and with L.B.J. and A.G.M. led the definition and implementation of the cascading modeling scheme and its implementation by other Co-Is under both HST programs and other related work. She led and conducted the *HST* far-UV observations and closely interacted with LBJ on most aspects of the work, in particular with in-depth discussions of the general magnetospheric science including the need to resolve the effects of metallicity versus field strength from signatures in the data. She worked on several aspects of the publication including writing some sections.

A.G.M. also provided the photochemistry hydro code modeling and contributed to the data interpretation and paper writing. He wrote Methods section III.3

P.L. provided the lower atmosphere chemistry code modeling and contributed to the data interpretation and paper writing. He wrote Methods section III.2

D.K.S. contributed to the implementation of the HST observing program, 1D lower atmospheric modeling, interpretation, and contributed to writing of section III.1 of Methods and paper in general.

J.S.F. was responsible for the X-ray observations obtained for this project. He analyzed the X-ray data, modeled the EUV emission, and conducted its interpretation with discussion with G.E.B and L.B.J. He wrote section III.6.1 of Methods with contribution from L.B.J. and G.E.B.

O.C. applied his MHD code and solar and stellar wind expertise to model to HAT-P-11b. He wrote section III.5 of methods with contribution from L.B.J.

T.K. provided the 3D GCM models and contributed to the writing of section III.1 of methods.

G.W.H. obtained the optical monitoring of the star activity that was used in two PanCET publications and contributed to the paper writing.

L.B., T.M.E., and H.R.W. contributed to the paper writing.

MLM. contributed to the obtention of the HST time and the preparation of the HST observations to collect the data presented in the manuscript. She contributed to the paper writing.

All authors contributed at different levels to the definition of the PanCET concept that is applied here for the first time.

**Competing interests**. The authors declare no competing interests.

# Extended Data Figures

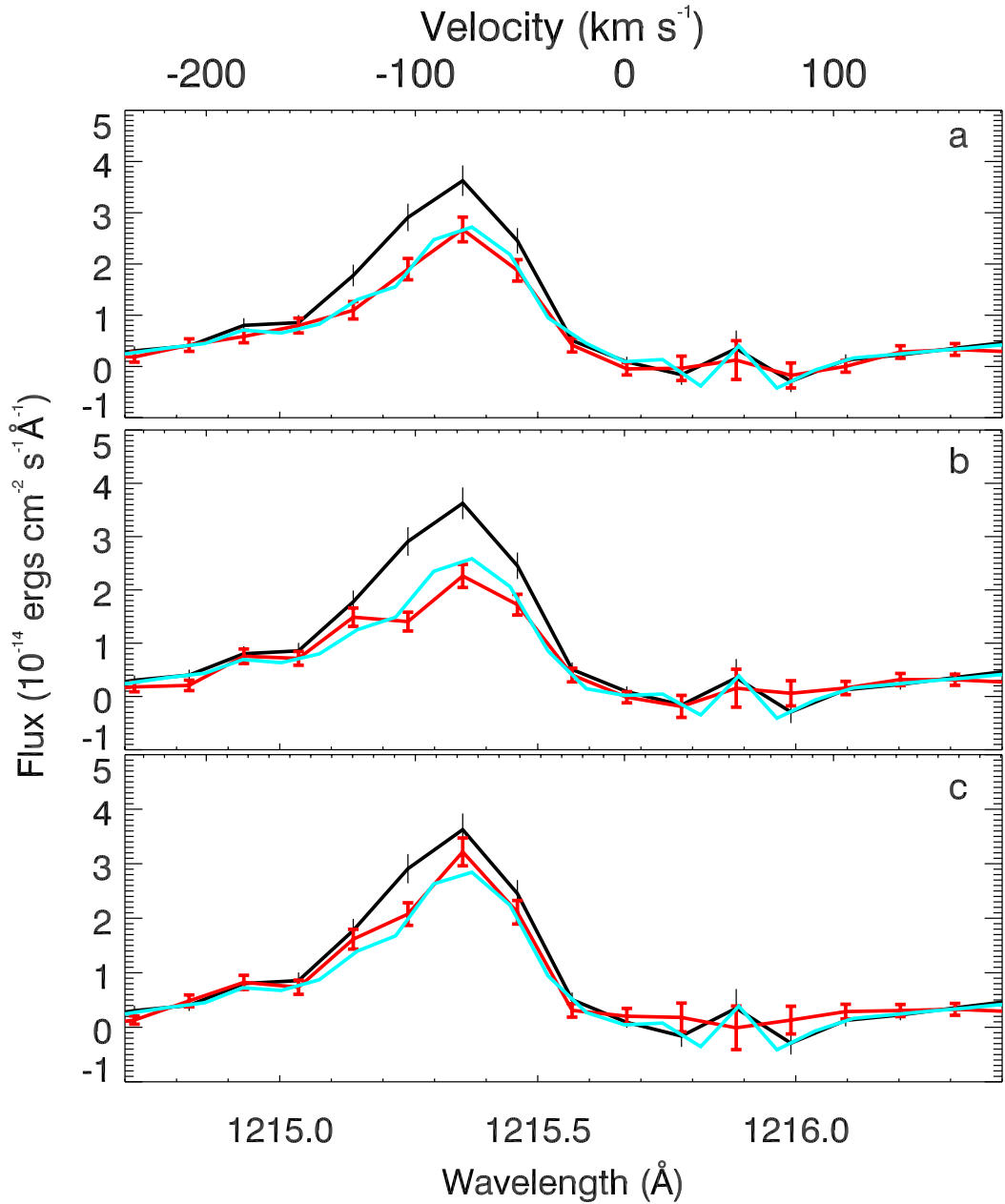

**Extended Data Figure 1**: **Lyman alpha model fit**. One of our best models fit (B=2.4 G, 2.35 x solar metallicity) compared to HAT-P-11b Lyα line profiles observed at selected phases of the transit event (HST visit 1 & 2 averaged). We show the out-of-transit Lyα line profile (average of orbit 1 of the two visits, black), in-transit observed line profile (red), and model best fit for the selected phase (cyan). Error bars represent the 1σ statistical uncertainties. **a,** HST orbit 5. **b,** HST orbit 4. **c,** HST orbit 3 (see Fig. 2 main draft for details).

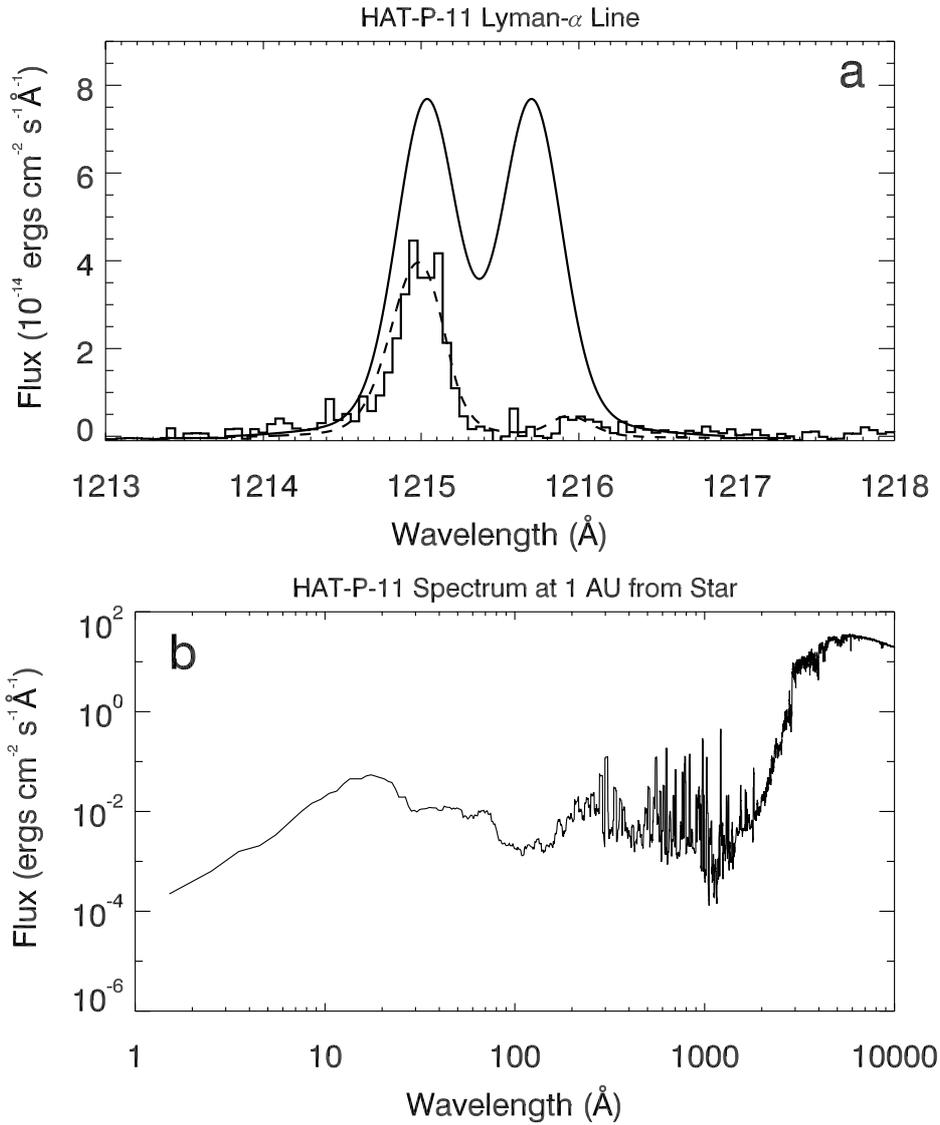

**Extended Data Figure 2: Stellar spectrum. a**, Reconstruction of the intrinsic profile (solid line) of the Lya line of HAT-P-11 using observations (histogram) and best fit model with an ISM [H I] ~ 4 x $10^{18}$ cm$^{-2}$ (dashed). **b,** HAT-P-11 full spectrum reconstructed at 1 AU from the star. The spectrum, in the range 1-54997 Å, is used as an input for all theoretical models developed in this comprehensive study. An ascii file of the spectrum is provided online.

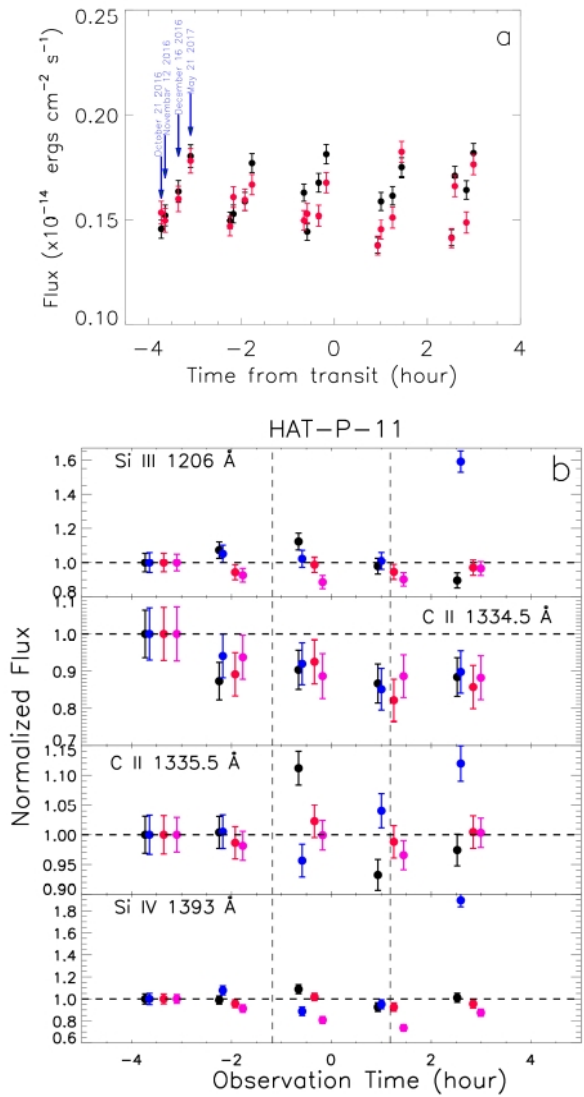

**Extended Data Figure 3: Light curves variability.** Transit light curves of HAT-P-11 versus time variability. **a,** Integrated flux of C II 133.45 nm (red) and C II 133.57 nm lines (black, scaled by the two lines' mean flux ratio ~1.41) versus time measured from the transit central time $T_c$. For clarity, dates of observations are only shown for the first HST orbit (T-Tc=-0.375), the other exposures being separated by a multiple of the HST orbit (1.5 hours). **b,** Normalized flux of FUV chromosphere lines. We notice a flare event during the fifth HST orbit of the transit observed on December 21 2016 (blue) in the SI III and Si IV lines, an activity that is not visible in the C II lines. An extended but weaker activity also appears at most orbital phases during the May 21 2017 transit (olive) for the Si III and Si IV lines but not for the C II lines.

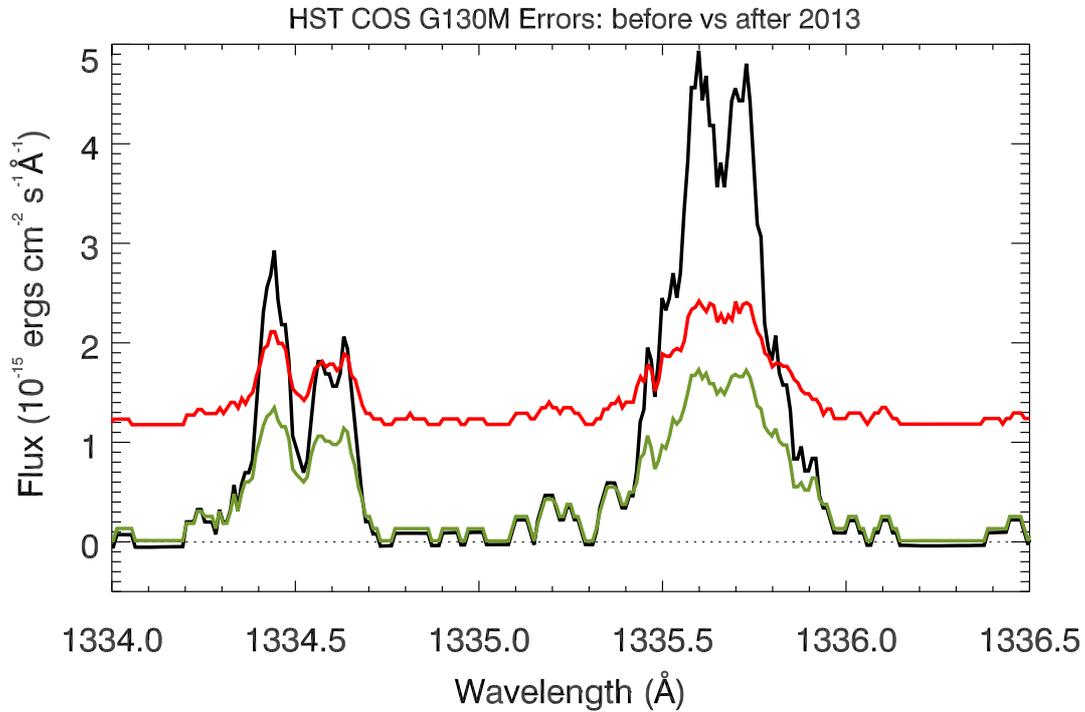

**Extended Data Figure 4: HST COS error.** HD 209458 exposure lb4m05knq obtained with COS G130M on Oct. 2, 2009 (Ballester & Ben-Jaffel, 2015)[38]. The stellar C II 1335 Å doublet spectrum (black) is compared to the statistical errors derived with the old (olive, e.g., CALCOS 2.14.4 or 2.18.5) and new (red, CALCOS 3.1.8) calibration pipelines. With the new pipeline errors (red), any detection of a transit absorption would be impossible. The new and highly inflated pipeline errors would also render the basic shape of the FUV spectrum of HD 209458 unmeasurable while other FUV dataset for this sunlike star are available (e.g., with *HST* STIS/G140L, STIS/Echelle data).

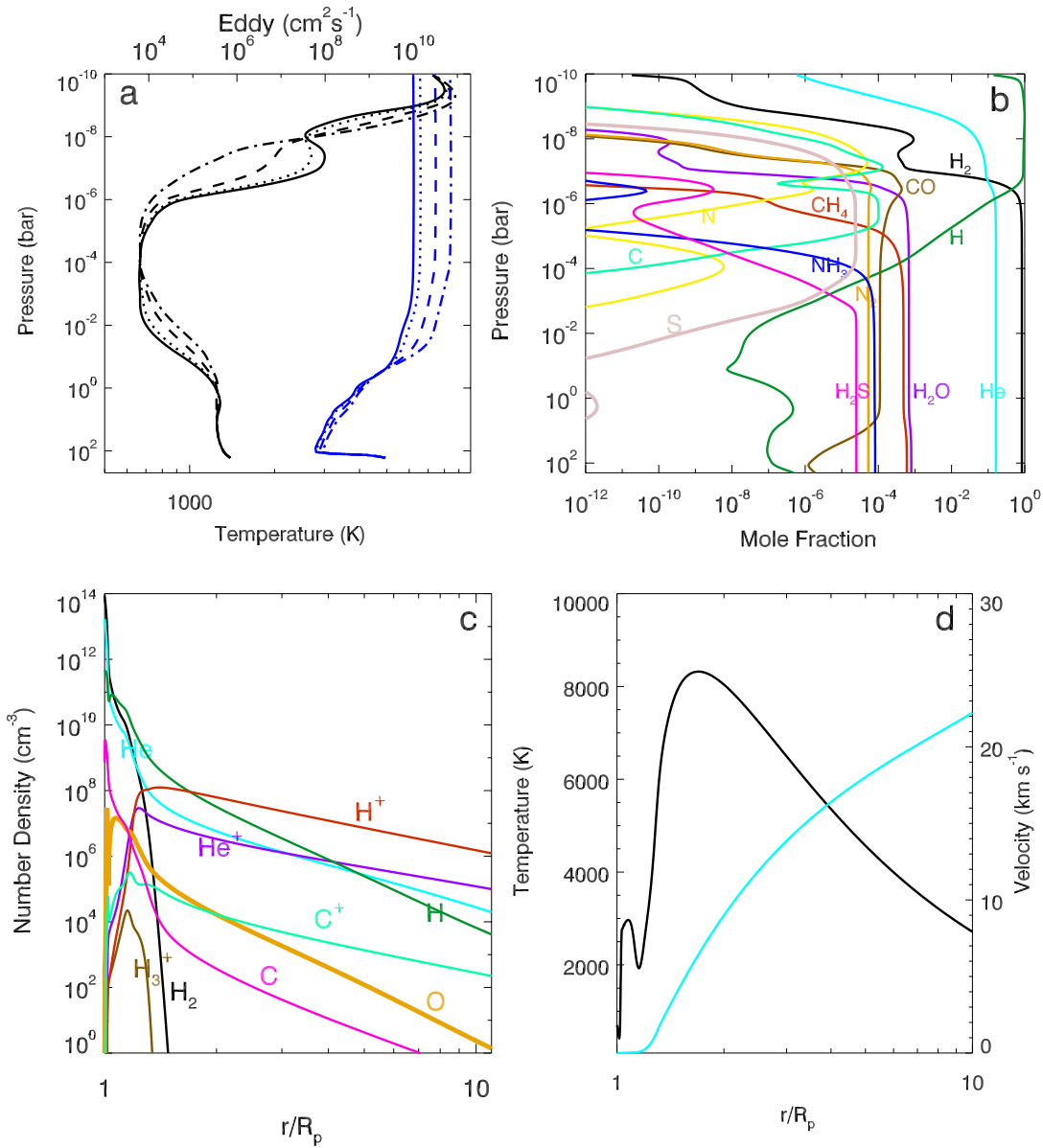

**Extended Data Figure 5: Middle and upper atmosphere models. a**, Atmospheric thermal structure (black lines) and eddy mixing (blue lines) for the atmosphere of HAT-P-11 b, under different assumptions of metallicity. The thermal profile is consistent with the conditions at lower atmosphere (section I) and upper atmosphere (section III). **b,** Model of species mole fraction distribution in the lower-middle atmosphere of HAT-P-11 b under the assumption of solar metallicity and thermal structure shown in **a**. **c,** Model of species distribution in the upper atmosphere of HAT-P-11 b based on mixing ratios displayed on **a** & **b**. **d,** Temperature and velocity distributions corresponding to upper atmosphere shown in **c**.

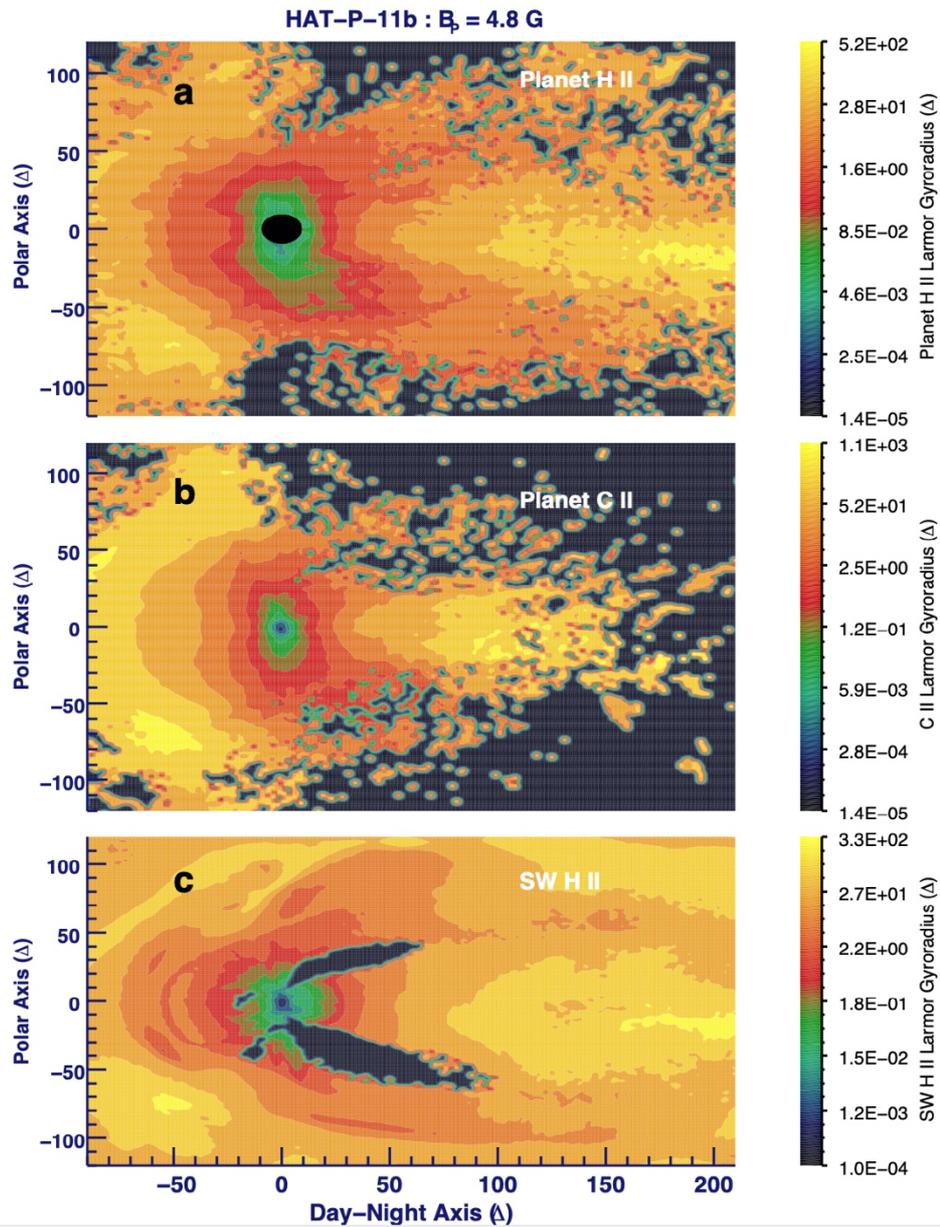

**Extended Data Figure 6: Plasma gyroradii.** 2D distribution of gyroradius of individual species in the noon-midnight plane. **a**, Planetary protons. **b**, Planetary C II. **c**, Stellar wind protons. For electrons (not shown), the gyroradius should be $m_i/m_e=100$ smaller. For the three plasma sources and for electrons, those scales are well resolved in our simulation, which allows us to properly describe charge separation and kinetic acceleration of species that are needed in the present study (see Supplementary Discussion X & Methods, section III.4 for more details).

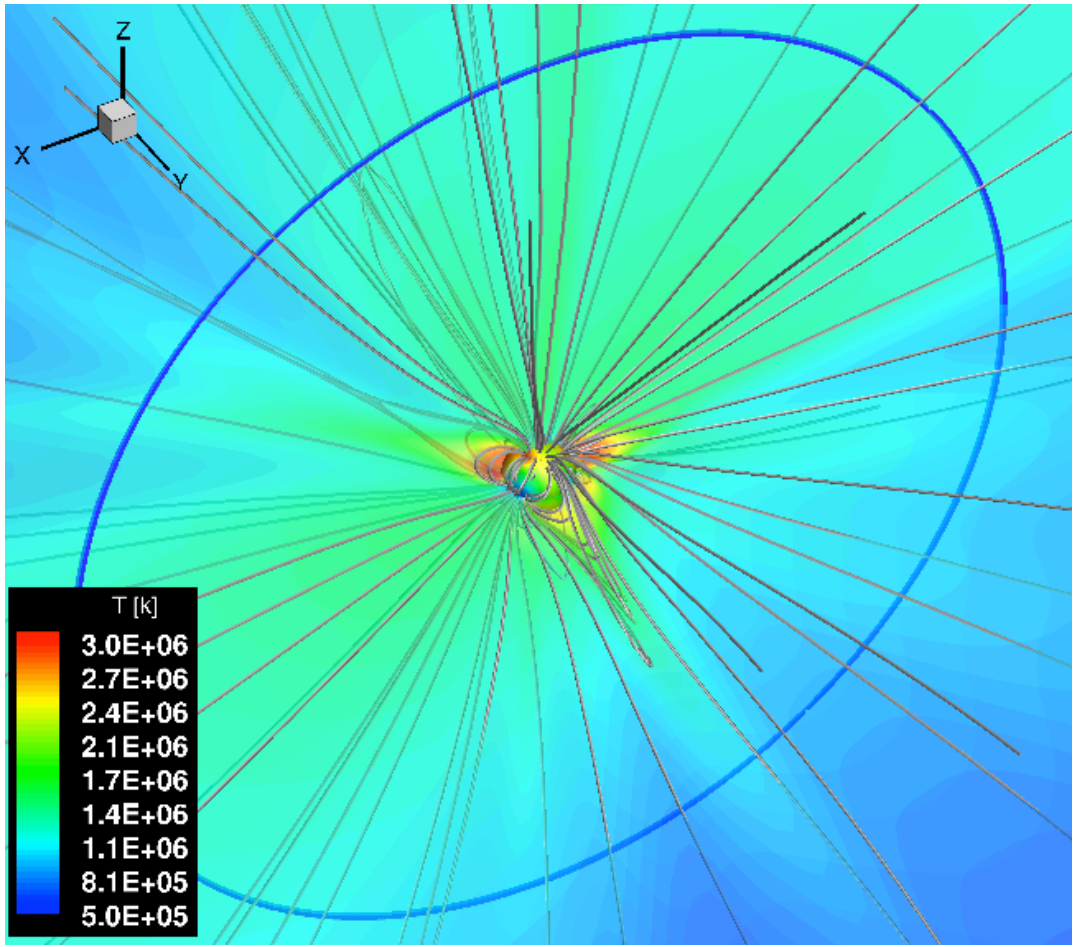

**Extended data Figure 7: MHD coronal model.** Temperature distribution in the exoplanet's orbital plane extracted from MHD 3D model simulation of HAT-P-11 wind (see Methods, section III.5 for more details).

# Supplementary Discussion and Tables: Signatures of Strong Magnetization and Metal-Poor Atmosphere for a Neptune-Size Exoplanet

Lotfi Ben-Jaffel[1,2*], Gilda E. Ballester[2], Antonio García Muñoz [3,4], Panayotis Lavvas[5], David K. Sing[6,7], Jorge Sanz-Forcada[8], Ofer Cohen[9], Tiffany Kataria[10], Gregory W. Henry[11], Lars Buchhave[12], Thomas Mikal-Evans[13], Hannah R. Wakeford[14], Mercedes López-Morales[15]

## Supplementary information related to Nature Astronomy Article : https://doi.org/10.1038/s41550-021-01505-x

[1]*Institut d'Astrophysique de Paris. Sorbonne Universités, UPMC & CNRS, 98 bis Bd. Arago, F-75014 Paris, France.*
[2]*Lunar and Planetary Laboratory, University of Arizona, 1541 E Univ. Blvd., Tucson, Arizona 85721, USA.*
[3]*Zentrum für Astronomie und Astrophysik, Technische Universität Berlin, D-10623 Berlin, Germany.*
[4]*AIM, CEA, CNRS, Université Paris-Saclay, Université de Paris, F-91191 Gif-sur-Yvette, France.*
[5]*Groupe de Spectroscopie Moléculaire et Atmosphérique, Université de Reims, Champagne-Ardenne, CNRS UMR 7331, France.*
[6]*Department of Earth & Planetary Sciences, John Hopkins University, Baltimore, MD, USA*
[7]*Department of Physics & Astronomy, John Hopkins University, Baltimore, MD, USA*
[8]*Centro de Astrobiología (CSIC-INTA), ESAC, P.O. Box 78, E-28691, Villanueva de la Cañada, Madrid, Spain.*
[9]*Lowell Center for Space Science and Technology, University of Massachusetts, Lowell, MA 01854, USA.*
[10]*NASA Jet Propulsion Laboratory, 4800 Oak Grove Drive, Pasadena, CA 91109, USA.*
[11]*Center of Excellence in Information Systems, Tennessee State University, Nashville, TN 37209, USA.*
[12]*DTU Space, National Space Institute, Technical University of Denmark, Elektrovej 328, DK-2800 Kgs. Lyngby, Denmark*
[13]*Kavli Institute for Astrophysics and Space Research, Massachusetts Institute of Technology, Cambridge, MA 02139, USA.*
[14]*School of Physics, University of Bristol, HH Wills Physics Laboratory, Tyndall Avenue, Bristol 1TL, UK.*
[15]*Center for Astrophysics | Harvard & Smithsonian, 60 Garden Street, Cambridge, MA 02138, USA.*
*email: bjaffel@iap.fr

### I. Grid for the magnetic field strength: reasoning.

PIC simulations are very time-consuming and the challenge in the present study was to define a strategy to decide a reasonable number of cases to scan the parameter space that defines the transit absorption. For HAT-P-11b, the phase of closest approach (periastron) is ~ 0.8745[13]. At that time, the exoplanet is ~ 12 $R_S$ from the star, while at greatest separation (apoastron), the distance is ~ 21 $R_S$. We compared the stellar wind ram pressure and speed from our MHD 3D simulations for these extremes' positions and for the mean transit position ~ 15.5 $R_S$. On average, the ram pressure is a factor 2 stronger for periastron and a factor 2 weaker for apastron compared to the time of transit. For the stellar wind speed, we found a 10% drop at closest approach and a 10% increase at the time of the greatest separation compared to the transit time. The derived ram pressure variation has little impact on the standoff distance MP of the magnetosphere (MP ~ $(P_{ram})^{1/6}$)[82]. For all speeds quoted above, the exoplanet is in the "day-side" orientation[83] all the time with a magnetospheric nose angle ~80° from the orbit. In summary, the wind's conditions over the HAT-P-11b eccentric orbit have no real impact on our calculation of the magnetic field strength.

For the stellar wind parameters described above, because the nose and tail of the magnetosphere are oriented approximately radial (i.e., along the line-of-sight), the projected area of the system during transit is nearly the size of the lobes of the magnetosphere across that direction. Both the C II or H I transit absorption require an object that must be at least ~10-20 $R_p$ extended across the line-of-sight in order to explain the absorption levels observed both one hour before and one hour after the optical ingress (e.g., Figure 2). To explain this result, we first approximate the absorption level as $(r_c/R_s)^2$ by an obscuring sphere of radius $r_c$ as a first guess. Then we use the orbital phase to get the planet's position relative to the edge of the stellar disk. If the projected orbital position is inside the stellar disk, the final size is $r_c$. But at ingress or egress (outside the stellar disk), one should add the separation between the planet's position and the edge of the stellar disk (particularly when that distance is larger than $r_c$). For the transit absorption so far derived in the range ~10 to ~30% at the orbital positions shown in Figure 2, an object of ~10-20 $R_p$ size is required. Thus, based on a scaling factor between dawn-dusk and dayside magnetopause sizes on Earth, one can translate the ~10 $R_P$ radius of the lobes to a stand-off distance of ~7 $R_p$ on the dayside of the nose of the magnetosphere. This stand-off distance requires a minimum exoplanet surface magnetic field strength of ~1.2 G for the stellar wind ram pressure assumed for HAT-P-11b (see Extended Data Table 3). For this reference case, our simulations do confirm a magnetopause stand-off distance of ~7 $R_p$ on the dayside and ~10 $R_p$ across the tail (e.g., Figure 2). For the range of stellar wind velocities derived from both our X-rays analysis and MHD simulation of the stellar corona, the magnetospheric nose/tail line should make an angle ~10° from the radial direction, if the spin and magnetic axis of the planet are coincident. When required by the analysis, we also consider the possibility of a tilt of the planetary magnetic field axis in order to check the sensitivity of our results to magnetospheric distortions (Main text).

This reasoning helped us define a grid of magnetic field strengths of 0.1, 1, 2, 4, and 8 times the ~1.2 G reference value for our final sensitivity study. As we find below, neither a weaker nor stronger magnetic field strengths are supported by the *HST* data.

## II. State of the art on observations of magnetotails in the solar system.

For reference, we briefly discuss here the state of the art of magnetotail measurements in our solar system. We focus our discussion on key aspects of three reference objects: Jupiter (like HAT-P-11b, with a strong magnetic field and under gravitational effects), Venus (no intrinsic magnetic field but with gravity), and comets (no intrinsic magnetic field and no gravity).

For Jupiter, many *in-situ* observations were obtained by Voyager 2, Galileo, and New Horizons (NH) at distances from ~50 to ~9000 $R_J$ tailward. All measurements tend to support a picture of a highly structured magnetotail, high fluxes, periodicity of a few to 25 days, and ion speeds in the range ~80 to ~700 km/s. Thermal plasma (low energy protons, high energy protons, and heavy 0.5kev low energy ions like O II) has been detected by Galileo[84] in the near tail with speeds in the range of 50 to 200 km/s in the region >~50 $R_J$ but at speeds 200 km/s anti-sunward at ~100 $R_J$, supporting the magnetospheric "wind" picture derived from energetic plasma (kev) detected by Voyager 2[85]. During the NH encounter with Jupiter at ~500 to 2500 $R_J$ downtail, magnetotail ions at ~90 to 190 km/s, ~140 to 760 km/s, and ~80 to 440 km/s respectively for O II, H II (protons), and $H_3^+$ have been estimated[21]. Interestingly, the ions' (protons, heavy species like O II) speed drops from ~430km/s (SW entry) to less than 200 km/s (within tail core) and then increases again to ~ 450-500 km/S (SW exit), as observed by Voyager 2/PLS during the spacecraft entry/exit of the core of the magnetotail at a distance ~ 6000 Rj tailward (Lepping et al., 1982, their Figure 4)[86]. All this evidence indicates the presence of relatively slow speed plasma inside the tail region that is distinct from the local SW speed outside the tail, a good indication that the tail dynamics are not yet controlled by the solar wind at 6000 Rj from the planet. In addition, the inter-comparison between the different regions so far explored by many spacecrafts shows that the plasma properties remain unchanged over large distances[84,86,87]. These observational facts remain unexplained by existing theoretical models[87].

For Venus (unmagnetized with gravity), all existing observations confirm that the tail plasma is flowing at the same solar wind (SW) speed at a distance of around 10-12 Rv and around a few thousands of Rv. Analysing Venera 9 & 10 and Pioneer Venus Orbiter data, Vaisberg et al. (2013)[88] reported ion speeds ranging from 45 km/s (O II) and 100 km/s (H II) at ~ 1$R_V$ to ~50 km/s (O II) and ~100-200 km/s (H II) at 4-5 $R_V$ but increasing to ~400 km/s for all species at 8-12 $R_V$. These findings were confirmed by Venus Express (Aspera-4 and MAG instruments) with O II ions detected at 10-12 $R_V$ with speeds ~ 400 km/s[89]. Finally, the SOHO/CELIAS mass spectrometer measured the plasma properties of the far Venus tail at a distance of ~ 7430 $R_V$ (4.5 $10^7$ km) downstream of Venus, showing relatively cold O II and C II ionospheric ions with speeds ~320 km/s consistent with the local solar wind speed[90]. It is thus apparent that the plasma flow speed in the Venus magnetotail (at distances >12 $R_V$) is comparable to the local solar wind flow speed.

For comets (unmagnetized & no gravity), during the Giotto encounter with comet Halley, the JPA and IMS instruments recorded a net radial velocity gradient ranging from a few km/s to ~200-260 km/ for distances between ~1300 km and 230,000 km from the cometary nucleus for protons and cometary ions (water ions)[91]. For the far tail region, the encounter of the Ulysses spacecraft with comet Hyakutake occurred at ~3.8 AU from the nucleus, allowing the detection of cometary ions (O II, C II, etc) with a plasma velocity shear between the surrounding solar wind (~750 km/s) and the plasma flow in the tail (~740 km/s) that had effectively disappeared[92]. In contrast, during the Ulysses encounter with comet McNaught at a distance ~1.6 AU downstream of the nucleus, the proton and He III speed dropped from ~780 km/S (SW) to 360 km/s (tail), which led to the conclusion that the ion tail of comet McNaught (composed of O II, C II, etc.) had not yet come to equilibrium with the surrounding solar wind[93].

It is important to stress that low-energy ions (the so-called hidden population with energy less than tens of ev) of ionospheric origin that dominate most of the Earth's magnetosphere are the main source of mass loss to the solar wind[81,7]. This fact, plus the arguments discussed above, supports our claim that a relatively low-speed and low-energy plasma population of planetary origin should persist all along the magnetotail of magnetized planets in contrast to unmagnetized bodies for which the interaction with the surrounding SW is more efficient. Faster (few hundred of km/s) and less abundant energetic particles should also be present but cannot be "seen" by their absorption imprint on the relatively thin stellar lines (FWMH~ 60 km/s for C II).

Before concluding this section, a brief mention of the complex problem of plasma acceleration in the magnetotail will be helpful, although a full discussion is beyond the scope of our study[94]. Indeed, different ions can get either the same energy or the same velocity, depending on the process that accelerates them. For example, ambipolar electric fields along field lines or magnetic tension force with unmagnetized ions may produce species with the same energy. In contrast, K-H instability would provide the same velocity to the different species[95]. As the dominant mechanisms at play in the magnetotail depend on the plasma condition between the stellar wind and planetary sources and on the planetary and interplanetary magnetic field and their corresponding multiple configurations, our approach using the self-consistent PIC simulation of the plasma at the kinetic scale (including the gravity of the object) is well-suited to handle the complex magnetospheric processes.

We finally emphasize that the finite size of the magnetotail in our simulation box should not be a limitation because we clearly observe a steady state of the kinetic properties of the escaping plasma over few tens of planetary radii in the simulation box. We have some indications in our solar system to support this statement. Indeed, the Jovian magnetotail is the only one object for which *in situ* observations have been obtained over large distances[21,87]. For instance, the SWAP instrument onboard NH reported 3-4-day periodic structures in the far tail, with a mean period of 3.5 days that is very close to the orbital period of the Jovian satellite Europa[21,87]. The quasi-coincidence with Europa's orbital period, and the identification of an $H_3^+$ component from the Jovian ionosphere, all provide a direct link between the plasma detected in the far magnetotail and their sources at either the Jovian ionosphere or at Europa[21,87]. Another interesting outcome of the NH mission is that SWAP found no evidence of peculiar variation in the plasma properties all along the Jovian magnetotail between ~500 and 2500 $R_J$. This high level of coherence and the related dispersion found in the properties of the plasma of the Jovian magnetotail led the SWAP team to conclude that "in general, tail structures were already largely developed by ~600 $R_J$ down tail and no longer, on average, significantly accelerating/slowing or expanding/contracting at greater distances."[87,96]

The properties discussed above for the Jovian magnetotail support our assumption of using uniform properties derived from the near-tail PIC simulation to define the farther magnetotail properties (e.g., Table 1).

### III. Exploration of modelling results.

We assume metallicities for HAT-P-11b in the range of 1, 2, 6, 10, 30, 50, 100, and
150 x solar for our sensitivity study. We find that changing the metallicity in the deep atmosphere only slightly modifies the thermal structure and average composition between the heavy species over the whole atmosphere. In addition, our coupled modelling from deep to upper atmosphere predicts that the total H I column changes only slightly, yielding a fairly constant H I escape rate of $(1.2-0.9) \times 10^{11}$ g/s, thus making H I a poor indicator of the metallicity bulk composition of the deep atmosphere.  In contrast, our models predict that the C II and O I column densities scale almost linearly with the assumed metallicity (Extended Date Table 2).

We ran our multi-species PIC code for different strengths of the intrinsic planetary magnetic field, implementing our code capability of tracking several
50

plasma species simultaneously—namely, the stellar wind source (H II and electrons) and the ionospheric/planetary wind source (H II, C II, and electrons).

For the planetary wind (PW) source, the PIC simulations produce a plasmasphere confined inside the magnetospheric cavity, where the plasma is flowing both toward the star and toward the magnetotail, forming an elongated plasma cloud of planetary origin that is amenable to transit detection. This plasmasphere corotates with the conducting ionosphere of the planet (Figures 4,5).

In the tail region, as shown by its streamlines, the flow is away from the star and nearly parallel to the stellar wind flow direction. In both the north and south lobes, we find a polar wind escaping along open field lines toward the tail (e.g., Figure 4). Based on our PIC simulations, we can classify the magnetotail plasma in terms of three main populations: (1) PW species flowing across closed field lines, (2) polar wind particles flowing along open field lines, and (3) SW species that enter the system from the tail. Each population is composed of ions and electrons, and so the system is globally neutral. Population (1) is a low-density plasma that should not exist in MHD frozen-in conditions, yet it was predicted by observational evidence and simple theoretical arguments[63]; our PIC code confirms it and provides its key properties, including the significant finding that most of this sub-population is lost in the interplanetary medium, which should heat and slow down the stellar wind. Only population (2) is of interest here because it remains at relatively low energies and exits the system principally along the tail. This picture is consistent with the current understanding of a low-energy hidden plasma population recently identified throughout the Earth's magnetosphere[5] and previously detected at other planets[21]. Population (1) is born at lower latitudes where particles are trapped by a complex current network related to the equatorial sheet, field-aligned currents, and ring currents. This current system is responsible for the acceleration of particles that get hot (keV) and fast (few hundred km/s) at speeds not sensed during transit by the narrow C II lines. Spectrally wide lines like the far wings of the Lya line (or the Mg II and Fe II lines on ultra-hot Jupiters[5]) could be better sensors of those populations.

In contrast to protons, which may originate in both SW and PW, the C II ion is only of planetary origin (the stellar wind carbon is too highly ionized)[32,97], and connects directly with the bulk composition of the exoplanet. In contrast to H I, H II and OI that are very sensitive to charge-exchange between them and with other species, C II is not sensitive to charge exchange with other species[98,99]. In addition, electron-impact ionization and recombination are very slow in the energy regime of a few electron-volts of the low energy population considered in the magnetotail region[99], thus keeping the average global C II abundance unaffected by the magnetospheric processes, unless the particles are picked up by the stellar wind (for example, outside the cusps regions). Note that C II dominates the C population because C I is quickly photo-ionized.

### IV. Length and Flattening of the magnetotail

For the large-scale properties of the magnetotail that are not captured by our limited-size simulation box, we assume uniform values that are evaluated as the average properties (density, bulk velocity, velocity dispersion) of the near-tail plasma in the simulation box on the night side (e.g., Table 1). In general, the tail properties must be modulated over time by variability in the external stellar wind and in the internal injection source. This type of modulation has been detected *in situ* in the extended magnetotail of Jupiter by many spacecraft[21,87]. In the case of HAT-P-11b, the modulation is even stronger and faster because of its nearly polar orbit around the star that crosses regions of stellar wind of opposite polarity and variable ram pressure every half orbital period (e.g., Extended Data Figure 7). We predict that a continuous breathing effect should affect the size of the magnetosphere due to the stellar latitude-dependent ram pressure (compression (smaller size) / depression (larger size) every half orbit). Our MHD simulations predict the IMF conditions along the full exoplanet orbit and show that during the transit observations (± 2 hours from mid-transit), HAT-P-11b is crossing the stellar current sheet with the IMF O-Z component nearly zero. For this reason, we only consider zero IMF conditions, but we plan to extend our simulations to non-zero IMF in the future, particularly if new transit data are obtained over extended orbital phases. Despite this modulation, for the purpose of evaluating the absorption by a column of the projected tail volume along the line-of-sight, only the integrated column, average mean velocity, and related dispersion really matter, which justifies the approximation adopted here.

To derive the average plasma properties in the magnetotail used in step 3 (main text), we first need to define the spatial size of the magnetopause (MP) inside which the plasma is confined in the directions perpendicular to the main flow direction (which lies approximatively in the OX direction in our simulation box, unless the flow is tilted). Both at Earth and Jupiter, the MP size on the north-south OZ magnetic field axis is smaller (flattening effect) than on the dawn-dusk OY direction[100]. Here, for all planetary magnetic field strengths considered, the plasma 3D distribution calculated with the PIC simulations show that the MP boundary is well resolved in both the meridional and equatorial planes and that the dawn-dusk MP size is ~2 times the size in the north-south direction (flattening along OZ).

We propagated the tail properties derived for every magnetic field strength case, allowing for a free but small tilt of the whole system (plasmasphere + extended tail) from the line of sight (that could be required by a tilt of the intrinsic magnetic field of the planet, or a specific configuration of the stellar wind that deviates from the MHD wind model used here).

## V. Sensitivity to model assumptions & overall error

In this section, we address the importance of feedback between modules (described in the 'Comprehensive global modelling of exoplanetary atmospheres' section in the Methods) and how taking them into account impacts our conclusions. In addition, we assess the sensitivity of our results to model assumptions and evaluate the corresponding impact on the overall error. The arguments provided below reinforce the robustness of our results regarding the B strength of HAT-P-11b and its atmospheric metallicity.

### V.1 Feedbacks between models

We consider appropriate feedbacks between the different models. For the lower atmosphere, we used input for the thermal structure from 1D RT and GCM simulations that then generated the chemical structure. The 1D hydrodynamic model used these results (including the prescriptions of the volume mixing ratios of the chemical species) for the evaluation of the upper atmosphere, which we then fed back to the 1D chemistry model and re-run the simulation. This was done a couple of times to properly describe the reality of the lower/upper atmosphere transition because eddy mixing rapidly transitions into advection as the dominating transport mechanism, so that molecular diffusion (and separation by mass) is not important (see sections 'Lowe-middle atmosphere' and 'Upper-atmosphere aeronomy' in the Methods).

For the transition between the hydrodynamic and PIC models, we do not consider any feedback from the magnetosphere into the ionosphere/thermosphere (below the base of our lower boundary) because the effect is negligible compared to the driving XUV source. For reference, it is now well accepted that the magnetospheres of slow spin planets (Mercury, Earth, etc.) are powered by the impinging solar wind[101]. Only about ~1 % of the kinetic energy of the solar wind is transferred to the whole magnetosphere. For Earth, Tenfjord et al., (2013)[102] used multi-spacecrafts observations over the 1997-2010 period to derive that the solar wind-magnetosphere coupling lead to an average transfer of ~0.6 % of the kinetic energy of the SW to the magnetosphere for all geomagnetic storm activity. About ~ 35% of that energy input is dissipated through coupling with the ionosphere/upper-atmosphere[102] (joule heating, particles precipitation, winds, etc.).

HAT-P-11b (~4.8 days spin period and small orbital distance), is not a fast rotator, making the stellar wind kinetic energy the dominant source powering its magnetosphere, besides the ionospheric outflow. To estimate the total energy input to the magnetosphere, we use our MHD 3D model of the stellar wind to derive the kinetic energy flux input $E_{sw} \sim 458.5$ erg cm$^{-2}$s$^{-1}$ at the orbital position of the planet (e.g., Extended Table 3). We also use the size of the system as calculated by our PIC model (e.g., Supplementary section IV). Assuming a stellar wind-magnetosphere efficiency of ~1%, we derive a power input to the magnetosphere smaller than ~$1.2 \times 10^{16}$ W for all values of the dipole magnetic field strengths considered. Of that total power input, only about ~ 35% (~$4.2 \times 10^{15}$ W) should be dissipated into the ionosphere/atmosphere, assuming the same energy partition as on Earth magnetosphere. For comparison, the XUV input power input is ~$4.9 \times 10^{16}$ W. This means that the magnetospheric energy input into the upper atmosphere is less than ~10% of the XUV energy input. To reach that conclusion, we implicitly assumed that after it is injected in the polar caps, the auroral energy is redistributed all over the sphere, an assumption that is well justified on Earth by observations for all altitudes of heat injection, particularly when including neutrals winds that are observed reducing the heating by a factor (10.)[103]. For reference, substantial heating redistribution from the polar caps' high latitudes to the lower latitudes is also clearly observed in Saturn's thermosphere with the Cassini mission[104].

Our final conclusion is that the additional local heating from the stellar wind-magnetosphere coupling should not exceed 10-20% of the local heating by the XUV source, a level of uncertainty that is comparable to the reconstructed XUV spectrum itself. For that reason, neglecting the feedback from the PIC model into the upper atmosphere model is reasonable for this comprehensive modeling of the global planet-star system.

The prediction of our MHD model for the stellar corona has been validated against the temperature derived from the observed X-rays spectrum of the star (e.g., 'Stellar wind plasma and interplanetary magnetic field conditions at the orbital position of the exoplanet' section in the Methods). This gives us the opportunity to work with accurate and self-consistent energy sources, namely the stellar XUV flux and the stellar wind energy inputs that power the HAT-P-11b atmosphere and magnetosphere, respectively.

### V.2 Sensitivity to model assumptions

The modeling tools used are the state-of-the-art, and there is a record of publications supporting this. Each numerical code used in this study has been independently validated against observations from solar system studies and exoplanets (see references attached to every model description in section 'Comprehensive global modelling of exoplanetary atmospheres' in the Methods).

One potential limitation of our 1D atmospheric models is the description of the night side of the planet, where we assumed that ionospheric plasma is injected at the lower boundary of the PIC model at half the rate of the dayside to account for potential day-night atmospheric circulation. We also tested night injecting at 1/10 of the dayside rate and obtained similar results for the magnetotail structure and composition. This means that the magnetotail structure and composition is principally dictated by the atoms escaping from the dayside, which are more accurately modeled in our approach. We explain this result by the fact that magnetospheric processes (bouncing particles between north and south hemispheres and multiple drifts) rapidly fill the equatorial region, particularly around the current sheet and outside a region of 2-3 Rp sphere around the planet. We mention those effects in Fig.5, stressing that the transit diagnostic is not sensitive to the region inside 3 Rp sphere, because the corresponding transit absorption (~ 3.4%) is less than the error bars (~4%) of the HST observations (e.g., Supplementary section VIII).

The sensitivity of our results to the description of the charge exchange process between ionized protons and the planetary neutral HI atoms is another issue that needs discussion. Our assumption of using only primary and secondary HI populations should not affect our predicted transit absorption, particularly the velocity along a line of sight that is compared to observations. Our conclusion is based on the similar problem of the interaction of protons-neutral species in the heliosphere. The comparison between a full kinetic and a two HI populations models (as done here) leads to a few % difference between the line-of-sight velocities of the models[78] (Osterbart & Fahr, 1992 (their Fig. 9 & 10)). This result can be understood from the fact that a higher order population (two charge exchanges and above) will be very small compared to the first two populations with a negligible weight on the total line of sight velocity. However, as stated in section 'Lya transit interpretation and modelling' in the Methods, some uncertainties remain on the total HI column along a line of sight and the size of the corresponding HI tail, which require further investigation. That uncertainty on the size of the HI tail does not affect our conclusions regarding the atmospheric metallicity (derived from C II) and the B field strength (derived from the Doppler velocities).

With the feedback between models described above, a main strength of our methodology is that it builds a global approach that should prevail over the details of the individual models as far as the final predictions of the model are tested against multiple observations that we discuss below.

## VI. Exoplanet Magnetic field strength: current context and perspectives

We find best fits of 1-5 G for the exoplanet surface equatorial field, an order of magnitude stronger than the ~ 0.2 G surface average of Uranus and Neptune. Our icy giants rotate fast at ~17 hours compared to the 4.8878-day period of HAT-P-11b (for synchronous rotation). However, the fast rotation of our giant planets cannot alone explain the difference observed in the strength of their magnetic fields (one order of magnitude between Jupiter and Neptune).  In addition, the *Juno* mission confirmed that the magnetic field of Jupiter has a strong non-dipolar component in northern hemisphere and a fully dominant dipole in the southern hemisphere[105], which questions our current understanding of the interior physics of planets and of the processes that create their magnetic fields. For example, a metallicity increasing with depth from the surface of the planet to the core is suggested as one of the requirements needed to explain the Jovian gravity and magnetic field data, in addition to the need for at least four layers that form the Jovian interior. This possibility means that the metallicity measured in the outer layer of the planet's interior might be very different from that in the deep interior[106].

For HAT-P-11b, the metallicity of the deep atmosphere (outer envelope of the planet) and the surface equatorial field strength (1-5 G; comparable to Jupiter's) make this planet interior akin to the Jovian conditions.  Based on an extensive set of convection-driven dynamo models in a rotating spherical shell, Christensen and Aubert (2006)[107] derived scaling laws for planetary magnetic fields consistent with the magnetic field strength being controlled by the available power and not necessarily by a force balance. Gastine et al. (2014)[108] showed that their scaling law is weakly dependent on the spin rotation of the planet. Using the internal heat measured during the Voyager encounter with the planet[109], they derive a Jovian magnetic field strength consistent with magnetometer *in-situ* measurements made by different spacecraft missions (*Voyager, Galileo, Juno*).

In the following, we thus use the fact that HAT-P-11b has a Jovian like interior and possesses a dominant dipole magnetic field. We use equations 2 to 5 of Gastine et al. (2014)[108] to derive the following scaling law for the dynamo B field ($B_{dyn}$) of HAT-P-11b (indexed H) and Jupiter (indexed J):

$$B_{dyn,H} / B_{dyn,J} = (P_H/P_J)^{0.35} * (\Omega_H/\Omega_J)^{-0.05} * (\rho_H/\rho_J) * (R_H/R_J)^{0.3} (r_{c,H}/r_{c,J})^{0.3} \quad \textbf{(Eq. 1)}$$

where $P_X$ is the convective power per unit mass of the planet X, $\Omega_X$ is its rotation rate, $\rho_X$ is its density, $R_X$ is its mean radius, and $r_{c,X}$ is the normalized radius of the top of dynamo region in the planet (0.83 to 0.9 for Jupiter). Using Eq. 3 of Gastine et al. (2014)[108] and assuming the two planets have the same relative internal structure, given their low metallicities, we derive the following scaling:

$$P_H/P_J = (R_H/R_J) * (F_H/F_J) \quad \textbf{(Eq. 2)}$$

where $F_X$ is the net heat flux of the planet. To connect the $B_{dyn}$ field with the field of the planetary surface, we use the fact that the dipole field strength falls as $1/R^3$ to derive[110]:

$$B_{equ,H} / B_{equ,J} = B_{dyn,H} / B_{dyn,J} * (r_{c,H}/r_{c,J})^3 \quad \textbf{(Eq. 3)}$$

Using Eqs. 1 to 3, and assuming the same normalized core radius $r_{c,H} = r_{c,J}$, we predict that HAT-P-11b should have a net internal heat flux of ~2.1 W m$^{-2}$ (compared to 5.5 W m$^{-2}$ measured for Jupiter) to produce the magnetic field strength of ~2.4 G derived here. The net intrinsic power from the planet should be ~1.9 x10$^{16}$ W (nearly 6% of the Jovian intrinsic power). This heating should enhance the thermal emission of HAT-P-11b with an equivalent intrinsic temperature excess of ~78 K. The secondary transit of the exoplanet was observed with *Kepler* at optical wavelengths but the night side was not detected[13]. With an upper limit of 2.x10$^{-4}$ on the thermal factor, the intrinsic power we derive here is consistent with that non-detection. Our prediction of the internal heat level of HAT-P-11b could be tested in the future by measuring the thermal emission of the exoplanet in the IR during secondary eclipse with a full phase curve. If that level of heating is confirmed, our findings should help define formation and evolution models for this type of exoplanets by providing key observables (metallicity, B strength), which opens new interesting prospects for the modelling of planetary interiors in the context of the diversity of thousands of exoplanets detected.

## VII. Metallicity

For HAT-P-11b we find best fits for a total metallicity of x1 solar (at 1σ) to x6 solar (at 3 σ) for the deep (200 bar) atmosphere our framework builds upon (Table 1). Our results disagree with the initial interpretation of the transit detection of water on HAT-P-11b in the 1.4-micron band by *HST*/WFC3, where the analysis yielded a high x 190 solar metallicity for the best fit, although within 1-sigma the results agreed with lower metallicities down to a few times solar (in the presence of high scattering clouds[14]). More recently, new PanCET observations of optical transits of HAT-P-11b made with *HST*/STIS and combined with the above *HST*/WFC3 near-IR data yield a best-fitting low metallicity of 0.11 x solar or 0.07-0.33 x solar at 1-sigma, and 2- and 3-sigma upper limits of 5 x and 85 x solar[15]. Thus, none of the most recent and independent analyses support that the planet formed and evolved to the high metallicities of our icy giants. Of significant note is that the two PanCET investigations rely on complementary chemicals, oxygen ($H_2O$) for optical/near-IR and carbon for the far-UV.

It is also interesting to note that other small mass exoplanets show a similar trend of low metallicity, despite the fact that these are closer to the size of Neptune. For instance, there is a water detection on HAT-P-26b that yields 5+22/-4 x solar O/H metallicity, and this is a lower mass (~ 0.06 $M_J$; 19 $M_E$) but highly inflated (~ 0.57 $R_J$) exoplanet[116,111] (more PanCET work is underway for this target). More recently, *HST* and *Spitzer* observations have detected water absorption and near-IR thermal emission on GJ 3470b yielding an oxygen O/H metallicity of 0.2-18 x solar[112], and GJ 3470b is even less massive (~0.04 $M_J$; 13 $M_E$) and inflated (~0.37 $R_J$). As the second most abundant element in the atmosphere, helium is also a good tracer to complement the diagnostic on the atmospheric composition (see Supplementary Section IX). For reference, our icy giants Uranus and Neptune have a total metallicity of ~ 70-140 x solar[113], quite different from the Neptune-mass exoplanets discussed above.

For the specific case of HAT-P-11b, in addition to its low metallicity, our finding of a strong magnetic field in the range 1-5 G moves the exoplanet further away from the Neptune family (~0.25 G for Neptune). Rather, HAT-P-11b is much closer to the Jupiter family, particularly with the magnetic field strength (~ 4.2 G) and low metallicity (2-3 x solar[114]) of Jupiter. In conclusion, HAT-P-11b would be the first of a family of exoplanets characterized by their small size, small mass, low metallicity, but strong magnetic field strength. *"Mini Jupiters"* would be an appropriate name for this family.

In terms of formation and evolutionary scenarios, the exoplanet metallicity should be compared with that of the parent star, which has twice the solar metallicity ([Fe/H]=0.31). It is curious that existing internal, evolutionary models for HAT-P-11b predict a 56x solar (or 28x stellar) metallicity[23], far from the low metallicity level derived in this study. It is likely that, much like the recent findings and questions regarding the low metallicity of Jupiter and its unusual magnetic field reveled by the Juno mission[105], the interior structure of these planets should be investigated not only in terms of size, mass, and metallicity, but also accounting for the formation of any convection-driven dynamo process. Properties such as metallicity and magnetic field strength will be much needed as constraints for those future models.

## VIII. Dipole magnetic field approximation: justification.

It is important to stress that considering either a Neptune- or Jovian-like magnetic field will not change our conclusions. Indeed, despite their potential dominance near the planetary surface, the field's quadrupole and higher order components drop much faster from the body center than the dipole component. Therefore, the latter effectively shapes the magnetospheric structure at distances of a few radii from the planet. We have at least two examples that clearly support this claim. First, Neptune's surface field has a quadrupole component that is comparable to the dipole one (strength ratio 1:1)[115]. Because the quadrupole strength drops approximately $1/r^4$ compared to $1/r^2$ for the dipole field, the latter dominates farther away than ~3-4 $R_N$ (strength ratio 0.1: 1), which led Mejnedrsen et al. (2016)[116] to consider only the dipole field for the MHD 3D simulation of Neptune's magnetosphere in their comparison to Voyager measurements. For reference, Mejnedrsen et al. (2016)[116] compared two simulation grids: the first with a size of 300x240x240 and a resolution of 0.5 $R_N$, and the second with a size of 500x400x400 and pixel resolution of 0.3 $R_N$ (both grids are comparable to our PIC simulation grid), finding the same physical properties from the two grids. Second, recent JUNO observations showed that the Jovian magnetic field has a complex topology with a strong non-dipolar component in the north hemisphere and a dipolar component in the south (strength ratio 3:1)[105], but at distances larger than 3-4 $R_J$, the strength ratio of the non-dipolar to the dipole components is notably smaller than ~ (0.3:1). For these reasons, during the last four decades the Jovian field was assumed to be dipolar[115]. Now, the key question is this: what is the impact of the opacity of a 3-4 $R_P$ inner sphere (where a potential non-dipolar component is acting) on the transit absorption compared to the whole magnetosphere? In the extreme case where the 3-4 $R_P$ wide region is assumed to be fully opaque, its transit absorption will not exceed 3.5% (~ $(\sqrt{10}*R_P/R_S)^2$), which is smaller than the error bars attached to both the H I or C II observations (e.g., Figure 2). Therefore, we deem that a dipole field is a reasonable assumption for our comprehensive study, and we see no practical reason to consider a Neptune- or JUNO-like complex B field.

## IX. Comparison with ground-based observations of extended He I on the planet.

He I transit absorption has been detected on HAT-P-11b at high spectral resolution with *Calar Alto*/CARMENES[117] (and confirmed with *HST*/WFC3 at low resolution[16]). The ~1.1% absorption was modeled with a cloud extending to beyond 5 $R_P$, and the low S/N ratio blue-shifted signature was fitted with an average anti-stellar motion of ~3 km/s. For instance, the He I absorption is produced by metastable He I atoms that typically result from the photo-ionization of He I atoms followed by the recombination of a population of ionized He II with electrons (the P-R mechanism) yielding newly created neutrals in an excited metastable state. Part of our cascade modeling approach already handles helium and other species, so we can extend the modeling to address the apparent smaller spatial extent in He I compared to C II and H I, as well as any blue-shifted signature. This effort is left for a future study, particularly tracking different species everywhere in the magnetosphere, and taking into account collisional processes that are not yet fully implemented in our PIC code.

## x. Adequacy of using PIC

The technical difficulties inherent to the huge contrast between plasma kinetic scales and the macroscopic scales of the magnetosphere have been extensively discussed in the literature[3,32,33,51,53,54]. For instance, a simple numerical calculation shows a huge contrast between kinetic scales like the species gyroradius (as low as ~0.1 km for H II in the Earth's inner magnetosphere) and a magnetosphere scale like the stand-off distance of the magnetopause[53,54] (~ 10 $R_E$~ 64,000 km). No existing machine, even using future exascale computers, can handle kinetic simulations that connect those scales, as it may take a few billion years for a single run[54].

The way to address the problem was to scale the plasma parameters in order to shrink the computing time while keeping most of the physics needed for the macro-system. For instance, for the Earth's magnetosphere, Buneman (1992)[118] proposed to use a large enough solar wind speed ($v_{sw}$~ 0.5 c, where c=0.5 is the speed of light). Another way commonly used to lower the computational cost is to reduce the ions-to-electron mass ratio $m_i/m_e$. For protons, $m_i/m_e$~ 1823 but most PIC simulations in the literature use a mass ratio ranging from $m_i/m_e$=4 (early work) to a few hundred. The reduced mass ratio makes PIC simulations feasible yet keeping a reasonable charge separation between species[56]. We also mention the scaling of the charge to mass ratio of species, which works if a large contrast is maintained between the scaled microscopic quantities and the large scale of the system[54]. In our study, we proceed as in most previous studies (reduced $m_i/m_e$ =100, electron charge to mass ratio $q_e/m_e$=0.6, and a stellar wind speed $v_{sw}$~0.22 as reference in the simulation). Further, it is important to stress that the decision on micro-scales scaling depends largely on the plasma macro-scales that must be achieved[31,53,54].

In the modeling of HAT-P-11b's magnetosphere, the only spatial constraint is that the simulation box must be large enough to cover the large structure of the magnetosphere and allows the ions to accelerate from speeds of a few km/s in the ionosphere to the average speeds revealed by the in-transit absorption Doppler shifts (~50 and ~100 km/s for C II and H I, respectively). Our simulation box has a regular grid 405x255x255 $\Delta^3$ (where $\Delta$= $R_P$/3 is the grid resolution) corresponding to a full size of the system of 135x85x85 $R_P^3$. Our grid size and corresponding pixel resolution are comparable to both MHD and PIC simulations of similar objects[53] (see Supplementary section VIII for more details). Figures 4c, d show that the PIC model recovers all the key structures (bow shock, magnetopause, lobes, current sheet, cusps). In addition, Figures 4a,b show that the acceleration region between the planet and the beginning of the magnetotail is roughly a sphere of ~60 $\Delta$ around the planet that is well resolved by the selected grid. We verified that the simulated ion speed follows a clear acceleration between the particles' injection at the planet's boundary and the magnetotail starting around ~60 $\Delta$ in the anti-stellar direction.

Another issue less covered in the literature is how PIC models handle waves propagation and interaction, particularly for a large-scale region like the magnetosphere. As any grid-based simulation, our PIC model resolves waves within the limitations imposed by the grid size (Debye length) and the time scale involved (plasma frequency, gyrofrequency, etc. of macro-particles). *Courant* condition and other conditions on the plasma parameters (Debye length, plasma frequency, etc.) efficiently reduce plasma instabilities[59]. Our PIC code allows the study of waves propagation, wave-wave, and wave-particles interactions, yet the simulation analysis is expensive because it requires one to record data at each time step. Indeed, the analysis is only possible as post-processing operations when performing the spatial transform of the selected variable (for example density) and saving the wanted harmonics at each step time[61]. Our assumption about non-reflecting boundary for the fields is efficient but leaves some residual backscattered waves (~1%) that can be a source of noise for the derived harmonics[62]. In this study, we do not see the benefit of such an effort and we leave it for future applications.

For reference, the presence of heavy minor species in a plasma requires that kinetic effects be included (like C II). With a heavier mass and a larger gyroradius, those species modify the plasma global properties[119]. MHD modeling of the plasma is not adequate as it would provide the same local speed for all species. One could consider multi-fluid MHD models but those are not self-consistent, particularly regarding the formation of polar outflow and the formation of the current sheet[120]. In both techniques, kinetic effects are dismissed, and the only way to explain the difference between averaged velocities of two species is to invoke different opacity distributions along a same line-of-sight.

The advantage of a PIC model is clear: it can describe different ion/atom velocities if the physics of the problem results in differing velocities, in addition to properly handling the opacity distribution along the line-of-sight. Our PIC model with scaled plasma properties, has the capability to separate the species mass and charges, and includes kinetic effects that properly describe the particles motions driven by Lorentz and gravity forces, self-consistently with the Maxwell equations that describe the electromagnetic fields. A PIC model is therefore the most appropriate and self-consistent way to describe the HAT-P-11b's magnetosphere, within the limitations (discussed below) imposed by the selected grid size and plasma time scales, and the handling of the boundary conditions for particles and fields in a finite simulation box.

## XI. Stellar spectrum reconstruction
### XI.1 Stellar XUV spectrum reconstruction

To model the spectral energy distribution (SED) in the XUV, we built a model of the emitting material in these layers, using X-ray spectra originated at the hottest temperatures and far-UV spectral lines formed at lower temperatures[70]. We adopted the photospheric abundance [Fe/H] = 0.3[10,12] for the corona, and solar photospheric relative abundances. The best fit was obtained using a two-temperature model: $\log T_1/T_2$ (K) $= 6.41^{+0.05}_{-0.03}$ / $7.06^{+0.22}_{-0.20}$, $\log EM_1/EM_2$ (cm$^{-3}$) $= 49.75^{+0.08}_{-0.10}$ / $48.79^{+0.27}_{-0.88}$, using an ISM absorption of $N_H \sim 5 \cdot 10^{18}$ cm$^{-2}$. The X-ray luminosity of the star is $L_x$ (erg s$^{-1}$) = 1.6 x $10^{27}$ (EPIC range, 0.3-10 keV), or 2.3 x $10^{27}$ (ROSAT range, 0.12-2.48 keV). We could construct an emission measure distribution ($\log T$ (K) = [4.0, 4.1, 4.2, 4.3, 4.4, 4.5, 4.6, 4.7, 4.8, 4.9, 5.0, 5.1, 5.2, 5.3, 5.4, 5.5, 5.6], $\log EM$ (cm$^{-3}$) = [50.50, 50.35, 50.20, 50.00, 49.80, 49.60, 49.30, 49.10, 49.10, 49.25, 49.30, 49.10, 48.95, 48.90, 48.70, 48.40, 48.00]) in the transition region using UV lines in our programs, assuming solar relative abundances to the adopted [Fe/H] abundance. We generated a synthetic SED in the range λλ 1 – 1200 Å using the calculated coronal (and transition region) model. We used the atomic database ATOMDB v3.0.9 in the X-ray fitting and generation of the SED. The derived extreme-UV (10-92 nm) luminosity is $L_{extreme-UV}$ (erg s$^{-1}$) ~ 4.01 x $10^{28}$ erg/s, and the X-ray part (0.5-10 nm) is $L_x$ ~ 2.36 x $10^{27}$ erg/s, which are consistent with early calculations[72]. The final XUV spectrum is shown in Extended Data Figure 2b.

Two recent publications[121] calculated the X-ray luminosity of HAT-P-11 and extrapolated the EUV luminosity based on empirical relations between the fluxes in the two bands. Their values differ from ours by up to a factor of 3 in X-rays, and a factor 2 in the EUV band[121]. Our coronal model calculates the EUV flux contribution more accurately than using scaling laws, because it uses both the information from X-rays and the HST/COS FUV observations. The overall error on the XUV flux should not exceed ~7.5%, based on the X-rays flux errors at the time of observation of the HST transits of HAT-P-11b, and on the fact that UV lines have lower errors in general.

### XI.2 Stellar FUV-IR spectrum reconstruction

The expected difference in the SED in the NUV between a K2V and a K4V star corresponds to two main components of the spectrum:

- the photospheric lines and continuum emission: this effect is taken into account using the PHOENIX spectra.
- chromospheric emission: this component depends more on the activity level (as indicated by their X-ray emission) than on the spectral type. For K type stars, NUV excess (above the photosphere continuum) is weakly correlated with the X-ray emission[122]. Based on that study, we decided to keep the NUV excess unchanged. Even accounting for the uncertain linear NUV-Xrays trend proposed by Richey-Yowell et al[122], the NUV flux used for HAT-P-11 would drop by 22% at most. Such a change in a relatively narrow spectral window would not affect the conclusions of our study because:

Photons between 140 and 300 nm are deposited at P >1 microbar for hydrogen-dominated atmospheres[28]. The eddy diffusion coefficient that was adopted in the lower atmosphere is moderately high and the advection velocity in the acceleration region of the upper atmosphere is also high. Fractionation by mass will be negligible in these conditions and we expect that the proportion of heavy and light gases in the upper atmosphere will be identical to the proportion in the lower atmosphere, regardless of the form in which they appear in the lower atmosphere. As a result, a moderately different NUV spectrum will not change the basics of the chemistry in the lower atmosphere.

**Supplementary Tables:**

| Instrument | Data Set name | Date Obs. | Time Obs. | Exposure time (s) | Orbital phase $T-T_C$ (hour) |
|---|---|---|---|---|---|
| STIS | od9m15010 | 2016-10-23 | 12:39:36 | 1917.06 | -4.042 |
| | od9m15020 | 2016-10-23 | 14:10:35 | 2178.17 | -2.489 |
| | od9m15030 | 2016-10-23 | 15:45:56 | 2178.16 | -0.900 |
| | od9m15040 | 2016-10-23 | 17:21:16 | 2178.17 | +0.688 |
| | od9m15050 | 2016-10-23 | 18:56:36 | 2178.17 | +2.277 |
| | od9e02010 | 2016-11-12 | 01:50:38 | 1900.12 | -4.108 |
| | od9e02020 | 2016-11-12 | 03:20:33 | 2178.08 | -2.571 |
| | od9e02030 | 2016-11-12 | 04:48:23 | 2178.16 | -1.107 |
| | od9e02040 | 2016-11-12 | 06:31:28 | 2178.19 | +0.610 |
| | od9e02050 | 2016-11-12 | 08:06:56 | 2178.16 | +2.201 |
| COS | ld9e01tnq | 2016-10-28 | 10:13:58 | 2297.18 | -3.728 |
| | ld9e01twq | 2016-10-28 | 11:36:30 | 3066.21 | -2.246 |
| | ld9e01urq | 2016-10-28 | 13:11:50 | 3066.17 | -0.657 |
| | ld9e01utq | 2016-10-28 | 14:47:11 | 3066.17 | +0.932 |
| | ld9e01uvq | 2016-10-28 | 16:22:31 | 3066.21 | +2.521 |
| | ld9m17d3q | 2016-12-16 | 07:29:10 | 1936.19 | -3.643 |
| | ld9m17d7q | 2016-12-16 | 08:50:48 | 2713.18 | -2.174 |
| | ld9m17d9q | 2016-12-16 | 10:26:08 | 2713.18 | -0.586 |
| | ld9m17dbq | 2016-12-16 | 12:01:2 | 2713.21 | +1.003 |
| | ld9m17ddq | 2016-12-16 | 13:36:48 | 2713.15 | +2.592 |
| | ld9m18oeq | 2016-12-21 | 05:04:54 | 1944.19 | -3.357 |
| | ld9m18ojq | 2016-12-21 | 06:24:20 | 2713.18 | -1.926 |
| | ld9m18osq | 2016-12-21 | 07:59:40 | 2713.18 | -0.338 |
| | ld9m18oxq | 2016-12-21 | 09:35:00 | 2713.21 | +1.251 |
| | ld9m18p2q | 2016-12-21 | 11:10:20 | 2713.18 | +2.840 |
| | ld9mh1a5q | 2017-05-21 | 18:07:01 | 1936.16 | -3.096 |
| | ld9mh1a7q | 2017-05-21 | 19:26:17 | 2713.15 | -1.773 |
| | ld9mh1a9q | 2017-05-21 | 21:02:11 | 2713.18 | -0.175 |
| | ld9mh1abq | 2017-05-21 | 22:39:29 | 2713.12 | +1.446 |
| | ld9mh1akq | 2017-05-22 | 00:12:16 | 2713.18 | +2.993 |

**Supplementary Table 1: Data log** of HST programs GO 14767 and GO 14625. TCT is transit central time was defined by propagating from zero phase on BJD 2 454957.8132067 and a period of 4.887802443 days (Hubert et al., 2017).

|  | Assumed volume mixing ratios at 10 µbar | | | | | Mass loss rate Hydro |
| --- | --- | --- | --- | --- | --- | --- |
| Metallicity | H | He | CO | $H_2O$ | $CH_4$ | $\dot{m}$ [g/s] |
| Solar x 1 | $4.149 \times 10^{-2}$ | $1.6 \times 10^{-1}$ | $1.306 \times 10^{-4}$ | $6.606 \times 10^{-4}$ | $1.715 \times 10^{-4}$ | $1.2 \times 10^{11}$ |
| Solar x 50 | $3.762 \times 10^{-3}$ | $1.628 \times 10^{-1}$ | $2.336 \times 10^{-2}$ | $9.99 \times 10^{-3}$ | $2.615 \times 10^{-3}$ | $9.3 \times 10^{10}$ |
| Solar x 100 | $1.929 \times 10^{-3}$ | $1.628 \times 10^{-1}$ | $1.165 \times 10^{-2}$ | $5.729 \times 10^{-2}$ | $1.733 \times 10^{-2}$ | $9.0 \times 10^{10}$ |
| Solar x 150 | $1.69 \times 10^{-3}$ | $1.612 \times 10^{-1}$ | $2.317 \times 10^{-2}$ | $7.666 \times 10^{-2}$ | $2.350 \times 10^{-2}$ | $8.9 \times 10^{10}$ |

**Supplementary Table 2:** The volume mixing ratio for the major constituent at the bottom boundary of the model domain, $H_2$, is calculated from $x_{H_2} + \sum_{i \neq H_2} x_i = 1$. The mass loss rate is calculated over the two hemispheres of the planet, i.e., $\dot{m} = 4\pi \sum u r^2$, where $\Sigma$ is the mass density, $u$ is velocity, and $r$ is the radial distance to the planet center at which both and $u$ are calculated.

| SW* coronal temperature $T_{sw}$ (K) | SW temperature at exoplanet orbit | SW speed $V_{sw}$ (km/s) | SW Ram Pressure $P_{ram}$ (N/m$^2$) | Sonic Mach $M_S$ | Alfvénic Mach $M_A$ | Magnetosonic Mach $M_{MS}$ |
| --- | --- | --- | --- | --- | --- | --- |
| $(2.6-3.7)\ 10^6$ | $(1.3-1.5)\ 10^6$ | 500-600 | $(1.4-2.0)\ 10^{-6}$ | 2.7-3.0 | 3.0-3.6 | 2.0-2.3 |

**Supplementary Table 3**: **HAT-P-11 stellar wind parameters** assumed in the present study. All Mach numbers indicate a super-magnetosonic flow. (*) From X-rays observations. All other values are based on the 3D MHD stellar wind model.

**Supplementary references.**